\RequirePackage{rotating}

\documentclass[11pt,a4paper,fleqn]{article}
\usepackage[utf8]{inputenc}
\usepackage{authblk}
\usepackage{geometry}
\usepackage{xcolor}
\usepackage[round]{natbib}
\usepackage{graphicx}
\usepackage{pdflscape}  
\usepackage{appendix}
\usepackage{threeparttable}
\usepackage{booktabs}
\usepackage{subcaption}
\usepackage{soul}

\usepackage{amssymb}
\usepackage{amsbsy}
\usepackage{bm}
\usepackage{amsmath}
\usepackage{amsthm}
\usepackage{graphicx}
\usepackage{mathtools}
\usepackage{setspace}
\usepackage{color, colortbl}
\definecolor{ashgrey}{rgb}{0.7, 0.75, 0.71}
\definecolor{columbiablue}{rgb}{0.61, 0.87, 1.0}
\definecolor{coral}{rgb}{1.0, 0.5, 0.31}

\definecolor{colBVAR}{HTML}{bababa}
\definecolor{colBART}{HTML}{d7191c}
\definecolor{colmixBART}{HTML}{fdae61}
\definecolor{colerrorBART}{HTML}{abd9e9}
\definecolor{colfullBART}{HTML}{2c7bb6}
\definecolor{plusgreen}{RGB}{26,76,57}

\definecolor{colcons}{HTML}{e31a1c}
\definecolor{colSV}{HTML}{a6cee3}
\definecolor{colhBART}{HTML}{1f78b4}

\usepackage{enumitem}
\newlist{steps}{enumerate}{1}
\setlist[steps,1]{label = Step \arabic*:}

\usepackage{adjustbox}
\usepackage{dcolumn}
\newcolumntype{d}[1]{D..{#1}} 

\definecolor{LightCyan}{rgb}{0.88,1,1}

\usepackage{caption}
\usepackage{subcaption}
\definecolor{nblue}{HTML}{000660}
\usepackage[colorlinks=true,urlcolor=nblue,linkcolor=nblue,citecolor=nblue]{hyperref}
\usepackage{subcaption}
\newcommand*{\myeqref}[2][Eq.~]{%
  \hyperref[{#2}]{#1(\ref*{#2})}%
}
\def\equationautorefname#1#2\null{%
  Eq.#1(#2\null)%
}

\begin{document}
\title{\textbf{Forecasting euro area inflation using a huge panel of survey expectations}\thanks{
 Huber and Pfarrhofer gratefully acknowledge financial support from the Austrian Science Fund (FWF, grant no. ZK 35).}}

\author[a]{Florian \textsc{Huber}}
\author[b]{Luca \textsc{Onorante}}
\author[a]{Michael  \textsc{Pfarrhofer}}
\affil[a]{\textit{University of Salzburg}}
\affil[b]{\textit{European Commission}}

\date{\today}

\maketitle\thispagestyle{empty}\normalsize\vspace*{-2em}\small\linespread{1.5}
\begin{center}
\begin{minipage}{0.8\textwidth}
\noindent\small \textbf{Abstract.} In this paper, we forecast euro area inflation and its main components using an econometric model which exploits a massive number of time series on survey expectations for the European Commission’s Business and Consumer Survey. To make estimation of such a huge model tractable, we use recent advances in computational statistics to carry out posterior simulation and inference. Our findings suggest that the inclusion of a wide range of firms and consumers' opinions about future economic developments  offers useful information to forecast prices and assess tail risks to inflation. These predictive improvements do not only arise from surveys related to expected inflation but also from other questions related to the general economic environment.  Finally, we find that firms' expectations about the future seem to have more predictive content than consumer expectations.
\\\\ 
\textbf{JEL}: C32, C53, E31, E47.

\textbf{KEYWORDS}: Tail forecasting, big data, Phillips curves, density forecasts.
\end{minipage}
\end{center}

\normalsize\newpage
\section{Introduction}
The Great Recession and the subsequent sovereign crisis have determined a long period of low inflation and put the monetary policy of the ECB to a severe test. Following the \textit{Great Moderation}, the link between inflation and economic slack seemed to return during the 2008 recession \citep{GiannoneetAl14} just to weaken afterwards, and inflation has usually proven challenging to forecast thereafter. Headline inflation partially recovered after 2017, mostly driven by energy prices, but core inflation
remained subdued,\footnote{We define core inflation  as HICP excluding energy and food.} despite the strong positive contribution of unconventional monetary policy as of 2014. In a further reversal, the Covid-19 crisis and the consequent supply side constraints, along with increasing energy prices, seem to have led to a rapid surge of inflation as of 2021.

The apparent instability of the link between inflation and other macro quantities makes inflation very difficult to forecast. \cite{LenzaJarocinski16} show the extent of this problem by adopting a comprehensive approach (focusing on several models of core inflation) and providing estimates of the output gap that best help in forecasting inflation. Their best model provides accurate forecasts but generates extreme values for the output gap. For instance, the model requires  that the output gap in the euro area (EA) has been about $-6$ percent on average in 2014 and 2015. Models that account for slower trend growth over recent years forecast core inflation worse. Interestingly, their results improve when they relate trend inflation to long-term inflation expectations. \cite{DelNegroetAl20} study the sources of this disconnect using VARs and an estimated DSGE model, finding a decrease in the  reaction of inflation to cost pressures and a shift in policy towards more forceful inflation stabilization as main determinants of the changes.

Modeling difficulties have translated to sizeable and systematic mistakes in actual forecasts of inflation. It is important to  note that these mistakes were not due to bad practices, but to the changing environment. This fact is documented, for example,  in  \cite{KontogeorgosLmbrias19}, who look in detail at the Eurosystem/ECB projection errors and show that despite their size and bias they fulfill criteria of optimality and rationality. A similar assessment by \cite{AlessietAl14} compares ECB and New York Federal Reserve forecasts after the Great Recession and reaches similar conclusions --- while suggesting that financial forward-looking data could have helped to reduce forecast errors.

When models based on observed quantities fail, either because of missing information or structural changes (e.g., a flattening of the Phillips curve), expectation-based forecasts emerge as a possible alternative, either to better inform structural models as in, e.g., \cite{DelNegroetAl20}, who find that the use of measured expectations produces more economically meaningful results than does the standard use of model-consistent rational expectations, or in reduced-form empirical analysis. The underlying idea is that, leveraging on the ``wisdom of the crowd'', the insertion of observed expectations may improve the understanding of inflation developments and ultimately lead to more precise inflation forecasts.

The natural candidates are inflation expectations. Inflation expectations are inherently latent and thus difficult to measure. They may be proxied empirically using either market-based or survey measures, both with advantages and drawbacks.  Most contributions focus on market-based measures of inflation expectations. These inflation expectations are considered an important indicator of the credibility of monetary policy. \citet{ciccarelli2021expectation} use inflation expectations derived from compensation markets and provide evidence of quantitatively important effects from US inflation expectations not only on domestic but also on international inflation. Specifically, using the EA as an example, they quantify an expectation channel that reinforces international spillovers from US inflation. Survey-based expectations are the second alternative. In principle they do not suffer from the same biases that characterize market-based expectations, but existing surveys generally represent the expectations of a small subset of the economic agents, focusing on specific groups of professional  forecasters that are well-informed and close to financial markets. This is the case of the most used inflation survey in the EA, the Survey of Professional Forecasters \citep[SPF, see, e.g.,][]{ECB_SPF, Banbura_etal21}.

Overall, the evidence about which measures of expectations work best is still not conclusive. Using Bayesian techniques, \cite{Moretti_etal19} estimate Phillips curves for the EA and find that expectations are the single most relevant determinant of price dynamics. They compare SPF-based expectations and measures of market-based inflation expectations (inflation linked swaps, or ILSs) at $1$, $2$ and $5$ year-ahead horizons, an find that all of them help explaining inflation in different subsamples. While inflation expectations seem to generally help inflation forecasting, they still present problems of built-in bias (risk aversion in the case of market-based expectations), timeliness of surveys (the SPF is conducted every three months), and generally of sampling and representativeness, since they are derived from financial markets or from surveying a few dozens of forecasters that are very close to financial markets.

Even for the purpose of inflation forecasting, there is a growing  consensus that surveys should be designed to more closely represent economic agents' perceptions, rather then perceptions of a small number of agents involved in financial markets, and  reflect more general economic conditions (as opposed to focusing on inflation exclusively). Additionally, whenever possible they should include perceptions about recent developments as well as intentions of future behaviour. For example, the ECB is currently  ending the pilot phase of the Consumer Expectations Survey \citep{ECBconsSurvey20},  with the scope of  collecting high-frequency information on perceptions and expectations of households in the EA, as well as on their economic and financial behaviour. The survey is conducted online each month, following a pilot phase targeting approximately 10,000 households, and is expected to contribute to the data used for policy analysis by the ECB. 

Well-known in the policymakers' arena, but relatively little used by the research community, the European Commission's Business and Consumer Survey has been providing highly representative, high-frequency expectations for the whole of Europe for about 60 years. Based on a wide pan-European sample currently including 40,000 consumers and 125,000 firms, the survey asks for perceptions of past developments and expectations for the future. Firms are asked about variables such as productions, order books, exports, selling prices and employment, or expected demand. Consumers are asked about their own economic and financial situation, about their perceptions of general economic and financial conditions, past and future prices, and intentions to make important purchases or to save.

This paper focuses on the explanatory power of this survey on inflation and its components, and highlights the role of expectations and their overall contribution to (short-term) inflation movements. We provide three main contributions:
\begin{enumerate}[label=(\arabic*),align=left,leftmargin=!,labelwidth=0.8cm]
    \item We extensively explore the survey. This survey is generally considered difficult to handle, as responses are qualitative. We compare sub-groups of survey respondents and assess their forecast accuracy. We assess the relevance of aggregated vs. disaggregated responses. We leverage on the whole set of responses that this Survey provides, rather than concentrating on inflation-related responses only. 
    \item Since the number of time series about expectations is huge, we develop an innovative econometric approach to deal with the curse of dimensionality.  Previous analysis of the survey, reported in the literature review, mostly considered one or a few answers at a time in the context of very simplistic models. Our approach includes subsets (or all of them) simultaneously and endogenously determines which parts of the survey predict inflation. 
    \item Besides considering headline HICP inflation, we extend our analysis to consider several inflation aggregates. We include core inflation, the domestic part of inflation which the central bank can best control via monetary policy; food; energy; non-energy industrial goods, and services. For each component we assess which sub-groups of survey respondents are important. Finally, we consider the FROOPP inflation rate, an indicator that Eurostat developed to better represent consumers' perceptions, and test whether consumers' expectations seem to be relevant for this aggregate as one would expect.
\end{enumerate}

We show that selected subgroups of the Survey improve the forecast of inflation compared to simple models containing the most important macro aggregates. Additionally, we show that the expectations of those directly involved in the trade of final goods (retailers and consumers) best identify signals of future inflation. The best results are observed at longer horizons and considering predictive likelihoods, suggesting that consumers and producers incorporate relevant longer-term information (both in the mean and variance) when forming expectations.

Other contributions in the literature have exploited this survey to forecast or understand the business cycle, and the Commission itself publishes European business cycle indicators (EBCI) based on it. Very few papers have explored the properties of this survey for nominal variables.\footnote{A notable exception is \cite{Arioli17}, which described the quantitative inflation survey. The authors conclude that there is still work to be done before using the quantitative expectations.} 
\cite{ClaveriaAl07} provide one of the first comprehensive analyses of the possible use of the Survey for forecasting. They analyse the possibility of improving forecasts for selected macroeconomic variables (including inflation) for the EA using the information provided by this survey. Considering several models and variables combinations, they conclude that information from the survey often yields improvements in point forecasts compared to the best model without survey information, but these gains are only significant in a limited number of cases. \cite{ClaveriaAl20} and \cite{ClaveriaAl21} further explore the information content of the Survey using genetic programming and spectral analysis. Compared to those, this paper provides a more comprehensive assessment of the information content of the Survey, based on an encompassing framework where the data are ``free to speak for themselves'' and are not hand-picked in any way.

\section{Data}

This section presents the dataset. Most of our regressions will explain a measure of inflation on the basis of two lags of the endogenous variable, the unemployment rate, industrial production (excl. construction), a money aggregate (M2) and up to two lags of the survey variables depending on the specification. Our dataset implies that each regression can have up to 1,400 explanatory variables. The following section explains how we deal with this huge number of regressors. 

\subsection{Inflation measures}\label{sec:inflationmeasures}
Our endogenous set of variables thus includes headline HICP inflation, its main components (i.e., food, energy, non-energy industrial goods, services) and core inflation.

In the EA, the Harmonised Index of Consumer Prices (HICP) is the most important measure of price dynamics. This is reflected in the inflation objective of the ECB, which is defined as an annual HICP inflation rate of 2\% over the medium term. Throughout this paper, we refer to inflation in the HICP as headline inflation. The HICP tracks the prices for a representative basket of goods and services for EA households. At the most disaggregated level currently available, Eurostat publishes 468 sub-indices for consumer prices. For policy purposes, some other higher level aggregates besides headline inflation are often considered by the ECB.

Core inflation is defined as the year-on-year percentage change in EA prices of ``all items excluding energy, food, alcohol and tobacco.'' This aggregate is important because food and energy are some of the most volatile components of the HICP. In addition, they are subject to substantial supply shocks, such as international price fluctuations of raw materials, environmental factors that can affect agricultural production, or fluctuations in the oil supply from the Organization of the Petroleum Exporting Countries (OPEC). All of these factors can heavily affect headline inflation, although they typically lie outside the reach of ECB policy. Movements in the volatile components may not be related to a trend change in the economy’s prices, but are more likely determined by temporary factors that may reverse themselves later. Thus, from a monetary policy perspective, changes in energy prices are not necessarily a sign of lasting inflation and, when they are included, can suggest spurious trends in aggregate prices. Given its implicit policy relevance, core inflation is also included in our econometric framework.

Finally, we consider the aggregate of Frequent out-of-pocket purchases, abbreviated as FROOPP, constitute a recently added special subaggregate of the harmonised index of consumer prices (HICP). FROOPP is calculated in the same way as other HICP special aggregates, using the data on prices and weights for components available to Eurostat. The aggregate is compiled to mainly represent purchases done by the consumers at least every month and paid for ``out-of-pocket.''\footnote{See also \href{https://ec.europa.eu/eurostat/statistics-explained/index.php?title=Glossary:Frequent_out-of-pocket_purchases_(FROOPP)}{ec.europa.eu/eurostat/statistics-explained/index.php?title=Glossary:Frequent\_out-of-pocket\_purchases\_(FROOPP)}.} Price increases for FROOPP have generally been higher than the headline inflation figure. These differences peaked in the period 2007-2008 when prices for many everyday products rose sharply, just to revert back in subsequent years. 

\subsection{The Business and Consumer Survey}
We now present the main features of the Business and Consumer Surveys. The Directorate General for Economic and Financial Affairs of the European Commission conducts regular harmonised surveys for different sectors of the economies in the European Union (EU) and in the applicant countries. They are addressed to representatives of industry (manufacturing), services, retail trade and construction sectors, and to consumers. These surveys allow for comparisons among different countries' business cycles and for monitoring the evolution of the EU and the EA economies, as well as monitoring developments in the applicant countries. High-frequency sampling, timeliness and continuous harmonisation are among their main qualities:
\begin{enumerate}[label=(\arabic*),align=left,leftmargin=!,labelwidth=0.8cm]
    \item The surveys are conducted on a monthly basis for manufacturing industry,  construction, consumers, retail trade, services, and financial services.  Additional questions are asked on a quarterly or semi-annual basis. Consumers, in addition, are asked about quantitative inflation expectations every three months. However, due to our need to have high-frequency data (and to additional problems in the quantitative expectations: see \citealp{Arioli17}) we will forecast inflation using the monthly data only.
    \item The surveys are conducted according to a common methodology, including harmonised questionnaires and a common timetable for the interviews. This ensures a high level of international comparability.
\end{enumerate}

The time sample covered is remarkably long for a European survey. The first survey covers the manufacturing sector as of 1962. Since then, the sector coverage was extended to the construction sector and to investment plans in the manufacturing sector in 1966, to consumers in 1972, to retail trade in 1984, and to the services sector in 1996. Since 2007, the Commission conducts a survey in the financial services sector at EU and EA level. The geographical coverage of the programme was progressively extended to include new Member States and new candidate countries. The programme currently covers all 27 EU Member States and five EU candidates. Survey results are then used by DG ECFIN for economic analysis, surveillance and short-term forecasting. Outside the Commission, the ECB, central banks, research institutes and financial institutions frequently use the EU survey data for both qualitative and quantitative analysis.

The sample size varies across countries according to the heterogeneity of their economies and to their population size. 134,000 firms and 32,000 consumers are currently surveyed every month across the EU. The nominal sample includes 38,000 units in the industry survey, 47,000 in the services survey, 28,000 in the retail trade survey, 22,000 in the construction survey, and 750 in the financial services. Due to non-response, the number of actually conducted interviews/filled-in questionnaires is usually about 30\% lower. The available raw answers are aggregated by individual question categories and provided by the European Commission as a huge number of time series. Table \ref{tab:variables} provides an overview of the total number of explanatory variables (labeled $M+K$ below) provided across country and survey subsets.

\begin{table*}[!t]
\begin{center}
\caption{Number of exogenous variables by survey sub-category (including lags).}\label{tab:variables}
\begin{threeparttable}
\begin{tabular*}{\textwidth}{@{\extracolsep{\fill}} lrrrrrrr}
\toprule
  \multicolumn{2}{c}{} & \multicolumn{5}{c}{\textbf{Classification}} & \multicolumn{1}{c}{} \\
  \cmidrule{3-7}
 & \textbf{Main} & \textbf{Building} & \textbf{Consumer} & \textbf{Industry} & \textbf{Retail} & \textbf{Services} & \textbf{All} \\ 
\midrule
  \textbf{EA} & 23 & 31 & 35 & 25 & 23 & 21 & 99 \\ 
  \textbf{Big 6} & 93 & 133 & 165 & 105 & 89 & 85 & 541 \\ 
  \textbf{Big 9} & 135 & 189 & 241 & 153 & 125 & 121 & 793 \\ 
  \textbf{Big 12} & 177 & 247 & 319 & 201 & 163 & 159 & 1053 \\ 
  \textbf{All} & 219 & 298 & 423 & 265 & 211 & 171 & 1332 \\ 
\bottomrule
\end{tabular*}
\begin{tablenotes}[para,flushleft]
{\footnotesize \textit{Notes}: Rows refer to country subsets, columns indicate different parts of the survey. ``EA'' are aggregate euro area measures; country subsets: ``Big 6'' (IT, FR, DE, NL, BE, ES), ``Big 9'' (Big 6 plus AT, IE, FI), ``Big 12'' (Big 9 plus PT, EL, SK). ``All'' refers to all available countries and survey indicators.}
\end{tablenotes}
\end{threeparttable}
\end{center}
\end{table*}

\section{Econometric framework}

\subsection{Big Data regressions using the singular value decomposition}
Our goal is to model inflation as a function of core predictors $\bm x_t$ of size $M$, and survey expectations $\bm z_t$ of size $K$. The main issue we face is that $K \gg T$, with $K$ reaching values up to 1,332 in our empirical work and $T$, which is the length of the training sample, being rather short. We solve this by using an efficient sampling algorithm coupled with a ridge-type shrinkage prior.

The point of departure is the following regression model which links $h$-step ahead inflation to $\bm x_t$ and $\bm z_t$:
\begin{equation}
\pi_{t+h} =  \bm x'_t \bm \beta + \bm z'_t \bm \gamma + \varepsilon_{t+h}, \quad \varepsilon_{t+h}\sim\mathcal{N}(0,\sigma^2),
\label{eq:baseregression}
\end{equation}
with $\bm \beta$ being a $M\times1$-dimensional vector of regression coefficients associated with the core predictors and $\bm \gamma$ is a huge dimensional vector of coefficients related to survey expectations. Moreover, $\varepsilon_{t+h}$ is a Gaussian shock with zero mean and variance $\sigma^2$. In terms of full-data matrices, this regression is given by:
\begin{equation}
\underbrace{\begin{pmatrix}
    \pi_{1+h} \\ \pi_{2+h} \\ \vdots \\ \pi_{T+h}
    \end{pmatrix}}_{\bm y}
    = \underbrace{\begin{pmatrix}
    \bm x'_1 \\ \bm x'_2\\ \vdots \\ \bm x'_T
    \end{pmatrix}}_{\bm X} \bm \beta +
    \underbrace{\begin{pmatrix}
    \bm z'_1 \\ \bm z'_2\\ \vdots \\ \bm z'_T
    \end{pmatrix}}_{\bm Z} \bm \gamma +
    \underbrace{\begin{pmatrix}
    \varepsilon_{1+h}\\ \varepsilon_{2+h} \\ \vdots \\ \varepsilon_{T+h}
    \end{pmatrix}}_{\bm \varepsilon} \label{eq: full}
\end{equation}
where $\bm y$ and and $\bm \varepsilon$ are $T \times 1$ vectors and  $\bm X$ and $\bm Z$ are $T \times M$ and $T \times K$ matrices, respectively.\footnote{Note that we transform all our variables prior to estimation to have zero mean and unit variance prior to estimation. Forecasts are subsequently transformed back to the original scale of the respective target variable.}

There are (at least) three issues with this regression model. First, notice that because $T \ll (K+M)$ overfitting issues arise. But these overfitting issues only materialize if we are able to compute the OLS estimator of $\bm \beta$ and $\bm \gamma$. Without further regularization this is not possible since  $\bm Z'\bm Z$ is not of full rank and thus not invertible. This constitutes the second issue. The third issue is that even after regularizing the precision matrix by adding $1/\delta \times \bm I_K$ to $\bm Z' \bm Z$ and then computing the covariance matrix through inversion, this might have to be repeated a large number of times (such as during an MCMC algorithm).

Several solutions exist for solving these issues. For instance, if the series in $\bm Z$ strongly co-move one could extract a small number of principal components (PCs) to soak up this information and replace $\bm Z$ in \autoref{eq: full} by its PCs. This approach has been advocated in, e.g., \cite{stock2002macroeconomic}, and often works well in practice. However, it could be that by relying on PCs we mask important idiosyncratic information across the elements in $\bm Z$. Hence, in this paper we offer a second solution based on a recent technique proposed in \cite{trippe2019lr}.

\subsection{Fast sampling using singular value decompositions}
In this section we focus on how to estimate $\bm \gamma$. Our approach is Bayesian and we will use a Horseshoe prior \citep{carvalho2010horseshoe} on $\bm \beta$, an inverse Gamma prior on $\sigma^2$ and a ridge-type Gaussian prior on $\bm \gamma$. These priors enable us to devise a Gibbs sampler  which iteratively samples from the full conditional posterior of $\bm \beta$, $p(\bm \beta | \bullet)$ with the $\bullet$ denoting that we condition on everything else. In fact, both $p(\bm \beta|\bullet)$ and $p(\sigma^2|\bullet)$ are available in closed form and textbook results \citep[see, e.g.,][]{koop2007bayesian} can be used. Sampling from the multivariate Gaussian conditional posterior of $\bm \beta$ is straightforward. The same holds true for the inverse Gamma posterior of $\sigma^2$.  It is the full conditional posterior of $\bm\gamma$ which gives rise to substantial computational challenges.

We use a singular value decomposition (SVD) to speed up sampling from $p(\bm \gamma|\bullet)$. This approach uses the SVD decomposition to decompose $\bm Z$ as follows:
\begin{equation*}
    \underbrace{\bm Z'}_{K \times T} = \underbrace{\bm S}_{K \times T} ~ \underbrace{\bm \Omega}_{T \times T} ~ \underbrace{\bm D'}_{T \times T},
\end{equation*}
with $\bm S$ and $\bm D$ being matrices with orthonormal rows and $\bm \Omega = \text{diag}(\omega_1, \dots, \omega_T)$ is a diagonal matrix with the eigenvalues $\omega_j$ along its main diagonal sorted in descending order.

We specify a conjugate prior on $\bm\gamma$:
\begin{equation}
    p(\bm \gamma | \sigma^2) \sim \mathcal{N}(\bm 0, \sigma^2 \delta \bm I_K).
\end{equation}
This ridge-type prior forces all elements in $\bm \gamma$ towards zero if $\delta$ is small. In many empirical applications (and if $K$ is large) this prior has been shown to work well \citep{griffin2013some}. In our case, we use a hyperprior $\delta^{-1} \sim \mathcal{G}(c_0, c_1)$ and estimate it alongside the remaining model parameters. We set  $c_0=3$ and $c_1=0.03$, leading to a prior that induces strong shrinkage.   Under this prior, the posterior of $\delta$ will be inverse Gamma distributed.

We use this rather simplistic prior structure to exploit substantial computational benefits. To illustrate this, notice that the conditional posterior of $\bm \gamma$ is given by:
\begin{equation*}
    \bm \gamma | \bullet \sim \mathcal{N}(\overline{\bm \gamma}, \sigma^2 \overline{\bm \Sigma}_\gamma),
\end{equation*}
with posterior moments given by:
\begin{align*}
    \overline{\bm \Sigma}_\gamma &= (\bm Z' \bm Z +  \delta^{-1} \bm I_K)^{-1}, \\
    \overline{\bm \gamma} &= \overline{\bm \Sigma}_\gamma \bm Z' (\bm y - \bm X \bm \beta).
\end{align*}
The key computational bottleneck stems from inverting $(\bm Z' \bm Z + \delta^{-1} \bm I_K)$ which is a huge matrix. Moreover, to sample from the Gaussian posterior one also needs to compute the Cholesky factor, a computationally intensive task that quickly becomes prohibitive for large $K$. 

To speed up the inversion of the posterior precision matrix, we can replace $\bm Z'$ with its SVD and use the Woodbury matrix identity:
\begin{equation}
    (\bm S \bm \Omega^2 \bm S' +  \delta^{-1} \bm I_K)^{-1} = \delta \left[\bm I_K - \bm S~ \text{diag}\left( \frac{\bm \omega^2}{\delta^{-1} \bm \iota_K + \bm \omega^2}\right)\bm S'\right], \label{eq:postcov}
\end{equation}
 with $\bm \Omega^2 = \bm \Omega' \bm \Omega$ and $\bm \omega^2 = (\omega_1^2, \dots, \omega_T^2)'$. The right hand side of \autoref{eq:postcov} is trivial to compute. The posterior mean $\overline{\bm \gamma}$ can be computed based on the SVD as well:
\begin{equation}
    \overline{\bm \Sigma}_\gamma \bm Z' (\bm y - \bm X \bm \beta) = \bm S ~ \text{diag}\left( \frac{\bm \omega}{\delta^{-1}\bm \iota_T + \bm \omega^2}\right)\bm D' (\bm y - \bm X \bm \beta).
\end{equation}
Especially if $K$ is huge (i.e., in the case of a few thousand covariates) naively computing the posterior mean also becomes computationally intensive and this representation avoids some of the demanding steps altogether. The covariance matrix in \autoref{eq:postcov} has a special structure but, unfortunately, is in general not sparse. Hence, computing the Cholesky factor becomes prohibitive and thus sampling from the Gaussian posterior will be difficult. Fortunately, there exist algorithms which allow us to avoid computing Cholesky factors altogether. The algorithm we use is based on \cite{cong2017fast}, which has recently been applied to economic data in \citet{hauzenberger2021fast}. 

To simulate from $\bm \gamma \sim \mathcal{N}(\overline{\bm \gamma}, \overline{\bm \Sigma}_\gamma)$ we rely on Algorithm 3 of \cite{cong2017fast}. This algorithm consists of two steps:
\begin{enumerate}[align=left,leftmargin=!,labelwidth=1.75cm]
    \item[\textsc{Step 1}:] We simulate $\bm a \sim \mathcal{N}(\bm 0_K, \delta \bm I_K)$ and $\bm b \sim \mathcal{N}(\bm 0_T, \text{diag}\left(\frac{\bm \omega^2}{\delta^{-1}\bm \iota_K+\bm \omega^2} \right)^{-1})$.
    \item[\textsc{Step 2}:] A valid draw from the posterior of $\bm \gamma$ is then obtained by computing:
    \begin{equation*}
        \bm \gamma = \overline{\bm \gamma} + \sigma \left[\bm a -  \bm S \left(\text{diag}\left(\frac{\bm \omega^2}{\delta^{-1}\bm \iota_K+\bm \omega^2} \right)^{-1}\right) \bm S' \bm a + \bm b\right]
    \end{equation*}.
\end{enumerate}
This algorithm has the advantage of scaling linearly in $K$  and is thus very fast. One caveat is that it requires using a ridge-type prior with a single shrinkage parameter on $\bm \gamma$, different to, e.g., global-local shrinkage priors. Note that a similar specification is used in \cite{giannone2021economic}. However, since we standardize our dataset prior to estimation, our approach is still capable of detecting relevant signals in $\bm \gamma$.

\section{Forecasting EA inflation}

\subsection{Design of the forecasting exercise and competing model specifications}
Each regression is of the form given by \autoref{eq:baseregression}, where $\bm x_t$ contains two lags of the endogenous variable, the unemployment rate, industrial production (excl. construction) and a money aggregate (M2). $\bm z_t$ contains two lags of the survey variables depending on the specification. We emphasize that the dimension of $\bm z_t$ is huge, as it can reach up to 1,300 variables including lags.

\begin{figure}
    \centering
    \includegraphics[width=\textwidth]{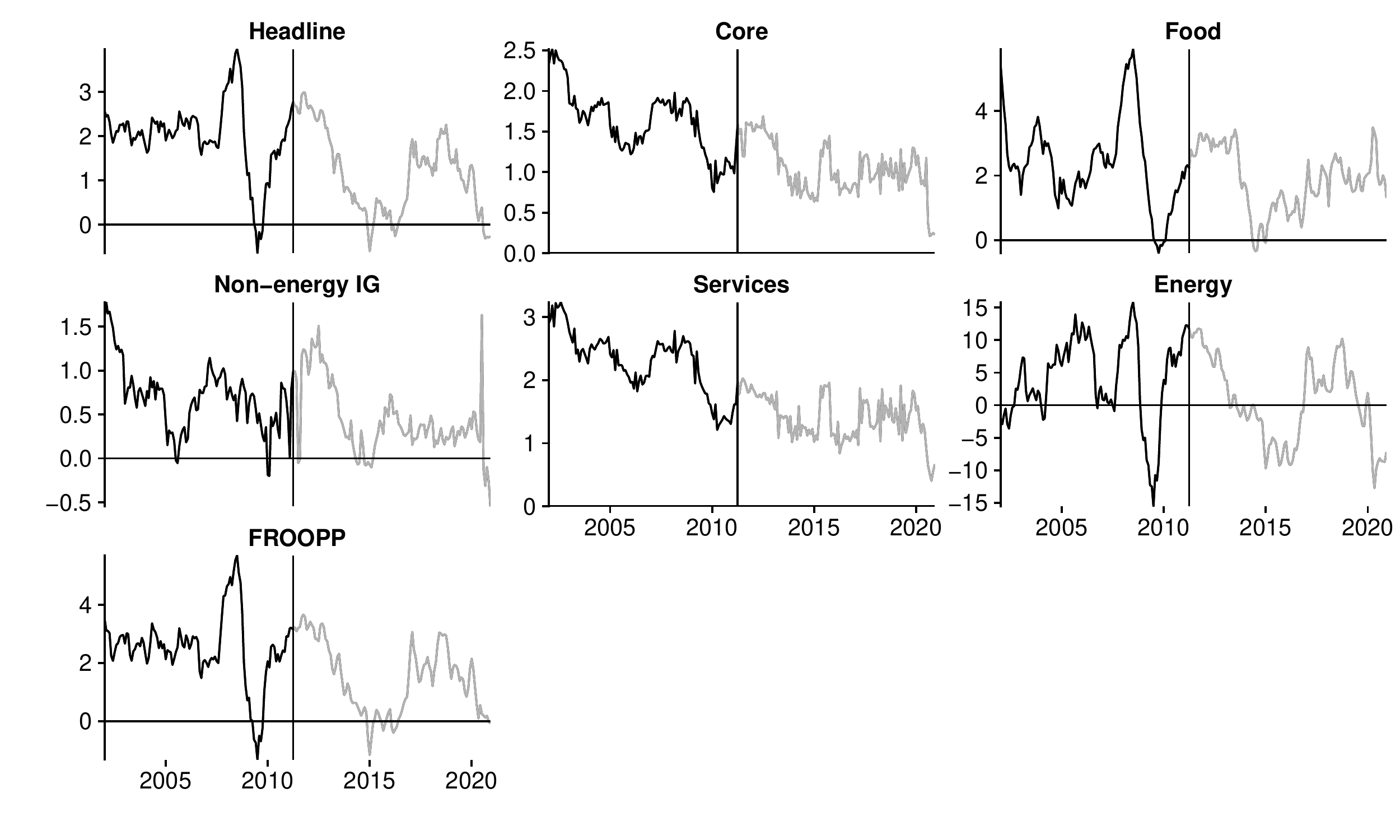}
    \caption{Inflation series. The vertical line indicates the start of the holdout period, such that the grey line marks the observations we use to evaluate our predictions.}
    \label{fig:data}
\end{figure}

We adopt a recursive forecasting design. Based on an initial training sample that ranges from 2002:01 to 2010:12 we compute one-month and one-quarter-ahead forecast distributions. After obtaining these we expand the estimation sample by one period until we reach the end of the hold-out sample (2020:12). A chart displaying our target variables is provided in Figure \ref{fig:data}, with the vertical line indicating the start of the holdout sample. For overall predictive performance, we assess the quality of the forecasts using two commonly used measures: the mean squared forecast error (MSE) for point forecasts and the log predictive likelihood (LPL) as a measure for density forecast accuracy due to its close relationship to the marginal likelihood \citep[see][]{geweke2010comparing}.

To zoom into specifics and investigate the origins of gains and losses in predictive accuracy in more detail, we also compute scoring rules which are capable of targeting distinct parts of the predictive distribution. In particular, we make use of the following scores. Let $\pi_{t+h}^{(r)}$ denote the realized value of inflation and $\pi_{\alpha,t+h}^{(f)}$ the $\alpha$-th quantile with $\alpha\in(0,1)$ of the predictive distribution. Following \citet{gneiting2011comparing} we define the quantile score (QS) as 
\begin{equation*}
    \text{QS}_{\alpha,t+h} = 2 \left(\mathbb{I}(\pi_{t+h}^{(r)}\leq\pi_{\alpha,t+h}^{(f)})-\alpha\right)(\pi_{\alpha,t+h}^{(f)}-\pi_{t+h}^{(r)}).
\end{equation*}
Note that the QS generalizes the absolute error to specific quantiles $\alpha$ (and yields the absolute error for $\alpha=0.5$). As a summary statistic for accuracy in different parts of the distribution, we rely on quantile-weighted continuous ranked probability scores (CRPS). They are constructed as follows:
\begin{equation*}
    \text{CRPS}_{i,t+h} = \int_{0}^{1}\text{QS}_{\alpha,t+h}\mathfrak{w}_i(\alpha)\text{d}\alpha~\approx~\frac{1}{J-1}\sum_{j=1}^{J-1} \text{QS}_{\alpha_j,t+h}\mathfrak{w}_i(\alpha_j),
\end{equation*}
which in its discrete version (for a grid of $j=1,\hdots,J$ quantiles with $\alpha_j = j/J$) corresponds to a weighted average of QSs based on the weights $\mathfrak{w}_i(\alpha)$ with $i\in\{\text{left},\text{right},\text{tails}\}$. We define the weights as $\mathfrak{w}_{\text{left}}(\alpha) = (1-\alpha)^2$, $\mathfrak{w}_{\text{right}}(\alpha) = \alpha^2$, $\mathfrak{w}_{\text{tails}}(\alpha) = (2\alpha-1)^2$, which target the left, right and both tails, respectively. In our empirical work we set $J=20$, and consider quantiles such that $\alpha\in\{0.05,0.1,\hdots,0.9,0.95\}$ in $0.05$ increments.

Our goal is to show how the addition of survey expectations affects forecasts. Hence, a natural benchmark of our model is  a regression that sets $\bm \gamma = \bm{0}_K$ and thus only exploits non-survey based information contained in $\bm X$. Another way of incorporating a large number of time series that measure expectations is to extract a set of principal components from $\bm Z$ and include these in $\bm X$. The resulting model is a factor-augmented predictive regression model in the spirit of \cite{stock2002macroeconomic}. Notice, however, that if expectations across countries and sectors do not feature a factor structure, including only a small number of principal components might mask important information. We will also consider this specification in several of our experiments.

\subsection{Summary of results}
Tables \ref{fig:MSEh3} to \ref{fig:expLPLh3} are heatmaps that present the results of our forecasting exercise. Each line of a table corresponds to an endogenous variable. Since we perform a large number of regression exercises, the table is structured in groups of columns. Within each group, we report the evaluation of the surveys on all the countries in our sample, for the Big-6 economies of the EA (DE, FR, IT, ES, NL, BE), the Big-9 (Big-6 with the addition of AT, IE and FI), and for the Big-12 (Big-9 plus PT, GR and SK). The last column of each block refers to the whole EU and to the EA (in aggregate terms). Note that the endogenous variable is always one of the indicated EA aggregates, while the country information refers exclusively to the surveys. 

The groups differ from each other because subsets of the survey are employed. For example, the first block only uses the main indices as regressors, the second group contains the questions that were posed to operators in the construction sector, and so on. The very last block features models that include all questions, no matter to whom they were posed. 

All reported forecast metrics are for three-months ahead forecasts and presented relative to the benchmark regression without survey information (i.e. with $\bm \gamma = \bm 0)$. These are computed by taking the ratio of a loss measure of a given model to the benchmark, subtracting 1 and multiplying by -1. This implies that the numbers can be interpreted as the percentage gains in predictive accuracy from including survey information.  Colors in the cells range from red (the surveys worsen the forecast) to blue (the surveys improve the forecast). A single dot besides the number indicates the best survey/country specification for each component. A dot within a circle marks the overall best performing model (in most cases, one of the SVD models is also the best performing specification overall, but this is not always the case).

Our focus on the three-months-ahead horizon is motivated by two reasons. First, in the very short term (nowcasting) shocks dominate, while at the three-months horizon the stochastic errors begin to average out. Second, most questions in the survey concern the following quarter, and the capacity of the surveys of spotting trends can be tested.  We also report one-step-ahead predictions in the appendix.

\subsubsection{Point forecasts}
We start our discussion by focusing on point forecasting accuracy first. Table \ref{fig:MSEh3} displays the three-step-ahead mean squared error for models featuring survey expectations relative to the non-survey information set. In the following, we discuss specifics for each of our target variables.

\begin{table}[t]
    \centering
    \includegraphics[width=\textwidth]{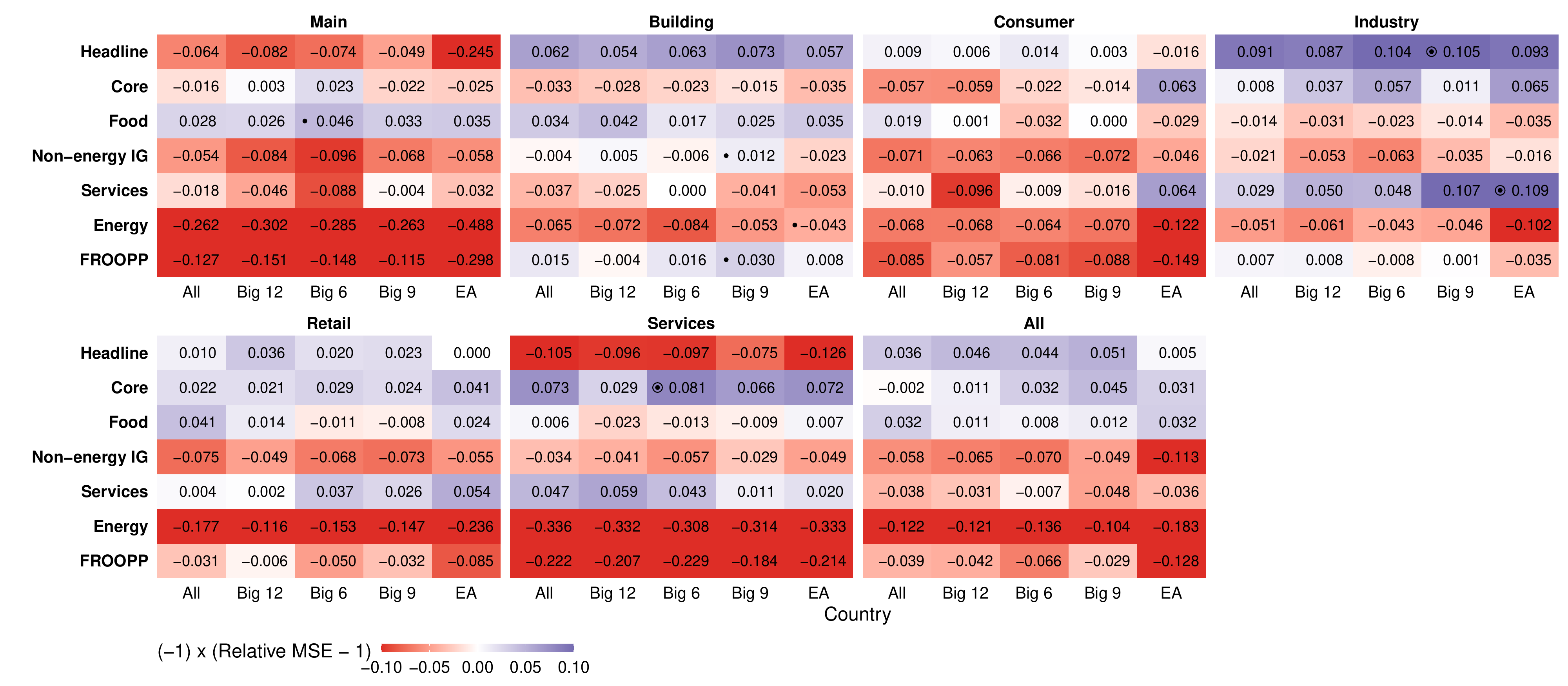}
    \caption{Three-step ahead mean squared error for models featuring survey expectations relative to the non-survey information set.}
    \label{fig:MSEh3}
\end{table}

We start by focusing on \textsc{Headline Inflation}. Being the aggregate of very heterogeneous components, headline inflation is notoriously difficult to forecast. The use of all questions (last block) signals modest improvements in point forecast accuracy, pointing towards adequate levels of shrinkage imposed by the ridge-type prior. Interestingly, it is possible to chose subsets of expectations that improve the forecast of headline inflation by a sizeable amount (about 10\%). In particular, expectations of the industrial sector, and to a smaller degree, construction, improve forecasts relative to the non-survey benchmark. Focusing on the country groups from which the surveys arise points towards smaller differences across forecasting models. This suggests that once surveys carried out in the big 6 EA economies are included, the additional contribution of adding more surveys from smaller countries diminishes.

Next we turn our attention to \textsc{Core Inflation}. As another aggregate that monetary policymakers watch closely, core inflation largely benefits form the same information as headline inflation.   In contrast to the results for headline inflation, however, inclusion of the building survey modestly worsens forecast accuracy, and services-surveys yield improvements. In fact, the latter results in decreases in MSEs of around 8\% while the former increases MSEs by 8\%. This may be explained by noting that the weight of inflation in prices of services in the construction of core inflation amounts to almost 60\%. Again, including all information from the surveys simultaneously (last block) signals moderate improvements and inclusion of surveys carried out in the big 6 countries yields the strongest gains.


Zooming into the components of inflation provides additional information on the determinants of the accuracy improvements for headline and core inflation. The first component that contributes substantially to the overall variation of headline inflation are \textsc{energy} prices. In line with an extensive literature on the topic, inflation in energy prices does not benefit at all from adding expectations. Energy prices in most countries depend on imported oil and gas, whose prices are notoriously difficult to forecast. Given that the information content is low, and despite our attempt to control the proliferation of parameters, noise dominates and predictive accuracy even decreases for point forecasts when adding survey information.

\textsc{Food} prices similarly adjust to dynamics largely determined by international markets, or by exogenous weather events. However, a significant share of overall food prices, processed food, has an important (domestic) EU component. The main aggregates and jointly including all survey series (see the block ``All'') modestly help forecasting food prices (with increases in predictive accuracy reaching almost 5\%).  Apart from these groups, our results suggest that including retailers surveys, who are at the end of the food chain production and whose policies largely determine final prices, has the potential to improve predictive performance. But this strongly depends on the country sample adopted. If we focus on all available countries at once, we find improvements of just over 4\% relative to the non-survey benchmark regression.

Point forecasts of inflation in \textsc{non-energy industrial goods} do not improve when considering survey information. Only when the building survey carried out in the big 9 countries is included we observe small but negligible improvements. 

 Forecasting inflation in prices for \textsc{services}, in turn, seems to largely benefit from data on expectations, as the inclusion of surveys generally does not hurt accuracy much but often improves our metrics by large margins. Industry surveys perform best in predicting prices of services (with improvements of up to 10\% lower MSEs). This suggests an interpretation based on prices being determined along the supply chain, with services often relying on upstream inputs. Retailers' information is another positive contributor to predictive accuracy, alongside services surveys themselves.

Finally,  an interesting question is whether consumers' expectations carry predictive power for the \textsc{FROOP}. The response is clearly negative. It appears that consumers, the group for which the index has been constructed, do not have clear information about the evolution of this index, even in the short term. The results for {FROOPP} provide empirical evidence to the supply chain hypothesis. Given that the forecast horizon is longer and industry produces a considerable part of domestic goods going to frequent purchases, the best information in relative terms (but improvements being negligible) can be found higher in the supply chain. 

Considering the predictive content across subsets of survey indicators, this first glance at MSE-based evidence suggests that the main aggregates and, surprisingly, consumer surveys do not improve the point forecast. By contrast, industry, building, retail and services surveys often improve predictive accuracy. For inflation (both headline and core) and some of the components (such as services and food), these improvements are sizable.

\subsubsection{Density forecasts}
Many professional forecasters and academics, only release/assess predictions of the conditional mean of inflation. Thus, they ignore predictive variances -- in other words, they do not provide information about how uncertain the forecast is. By contrast, we evaluate the forecast on the basis of the full predictive density in this section.

\begin{table}[t]
    \centering
    \includegraphics[width=\textwidth]{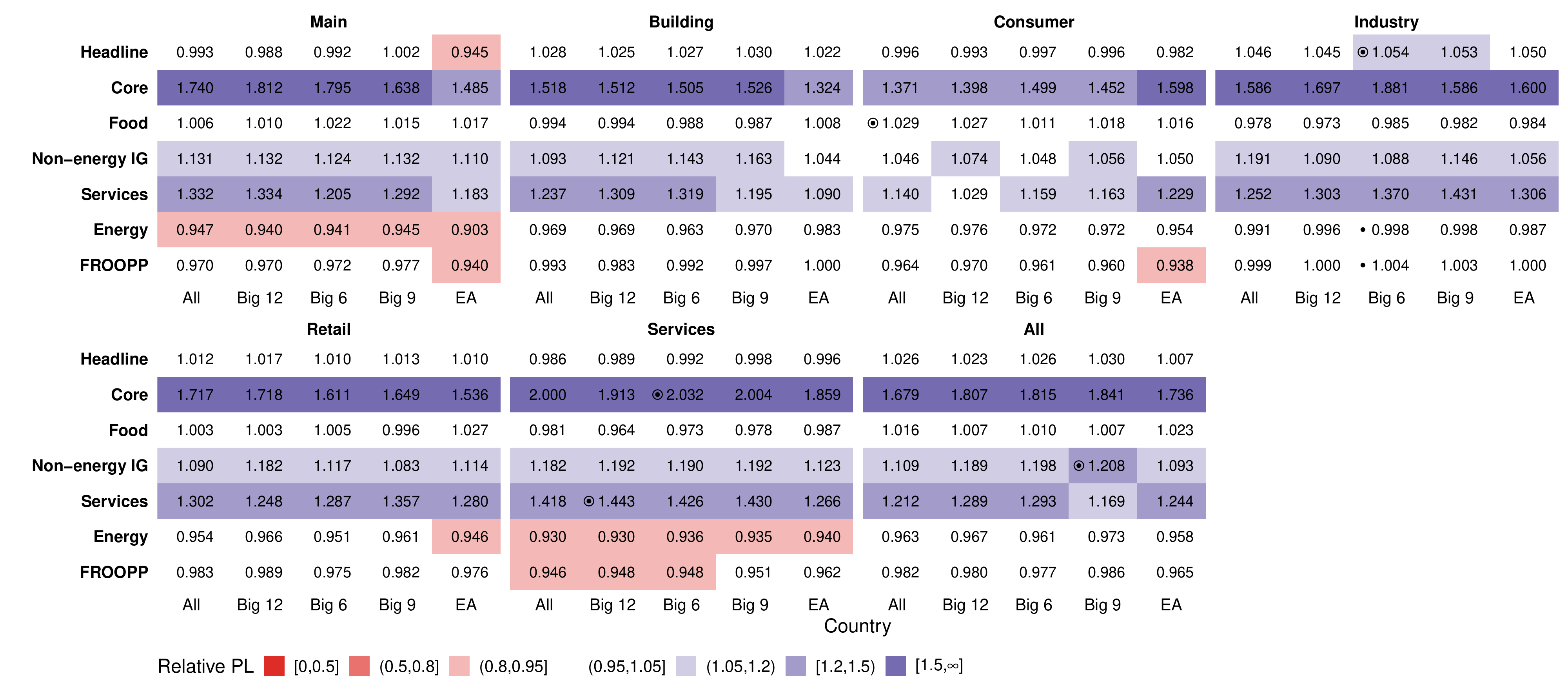}
    \caption{Three-step ahead predictive likelihood for models featuring survey expectations relative to the non-survey information set.}
    \label{fig:expLPLh3}
\end{table}

The previous section showed that including survey expectations often improves forecasts of the conditional mean of headline inflation and its components. Table \ref{fig:expLPLh3} reports predictive likelihoods (PLs) for three-month ahead forecasts. These  take higher-order moments of the predictive distribution into account. In the following, we provide an overview on headline inflation, core inflation and the individual components.

For headline inflation, different to improvements in terms of MSEs, the predictive likelihood does not increase much when considering surveys relative to the baseline regression. It is, however, worth mentioning that the best overall performing model for density forecasts (albeit at a small margin) is the one including industry surveys. The results for core inflation, by contrast, are likely to be particularly useful for central bankers. All survey-based specifications improve density forecasts to varying degrees. The improvements in relative predictive likelihoods are substantial, sometimes by a factor of two (especially when services surveys are included).

Investigating predictive accuracy in terms of PLs for the components of the HICP, we find much more homogeneous patterns when compared to MSEs. While there are virtually no improvements for energy prices or the FROOP aggregate, and only very limited gains for food, density forecasts of inflation in services profit across most information sets we consider. A key difference to point forecasts is that survey information pays off when the focus is on density predictions of non-energy industrial goods. This implies that survey information, apart from giving a sense of direction of future inflation, is also valuable in predicting its dispersion.

\subsection{Forecasting results over time}
In stable economic periods, very simplistic forecasting models often perform at their best. This is, however, when they are the least useful. A model that performs comparatively well during crises or periods of high variability and structural changes provides substantial value added for policy makers.

\begin{figure}[t]
\includegraphics[width=0.323333333333333\textwidth]{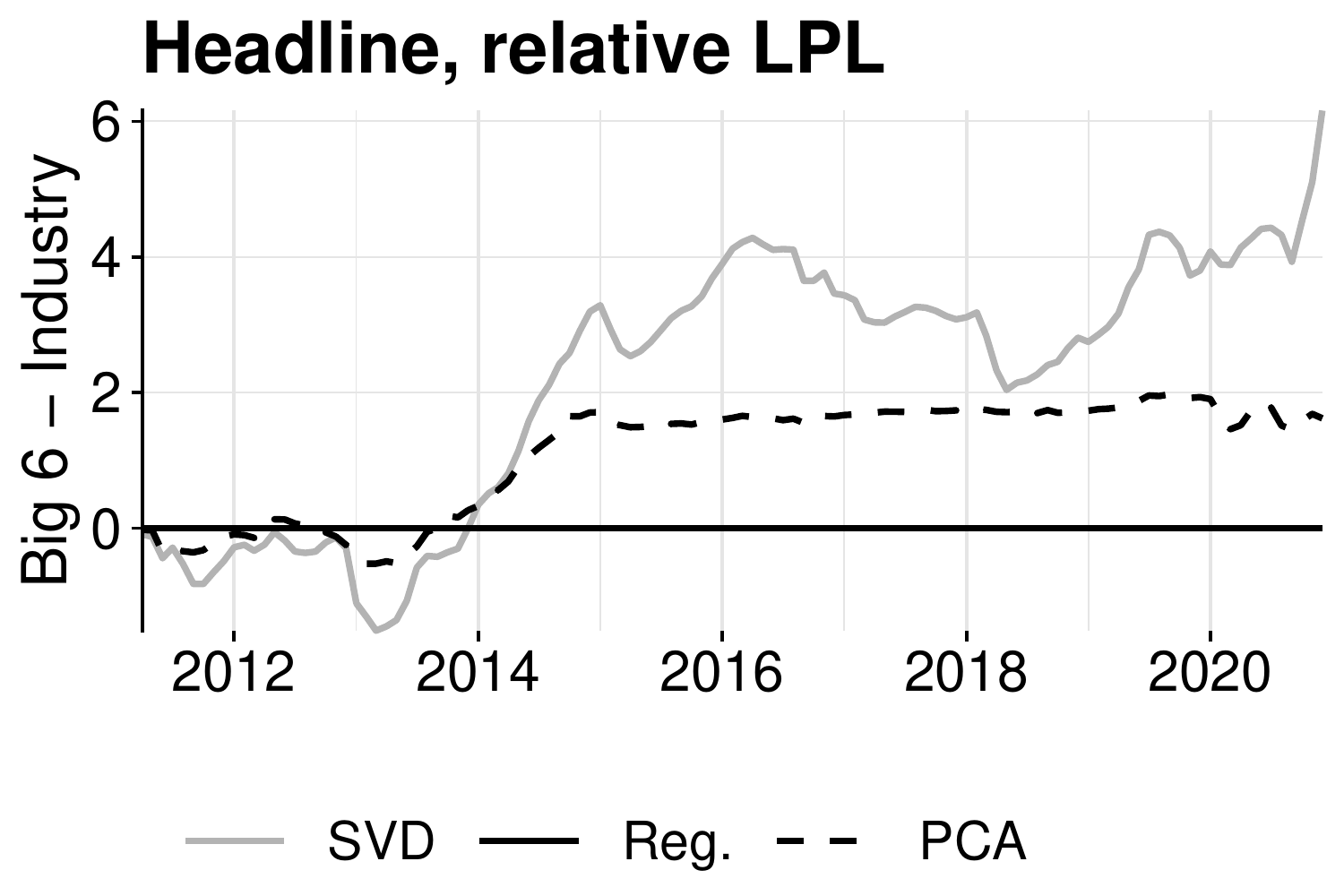}
\includegraphics[width=0.323333333333333\textwidth]{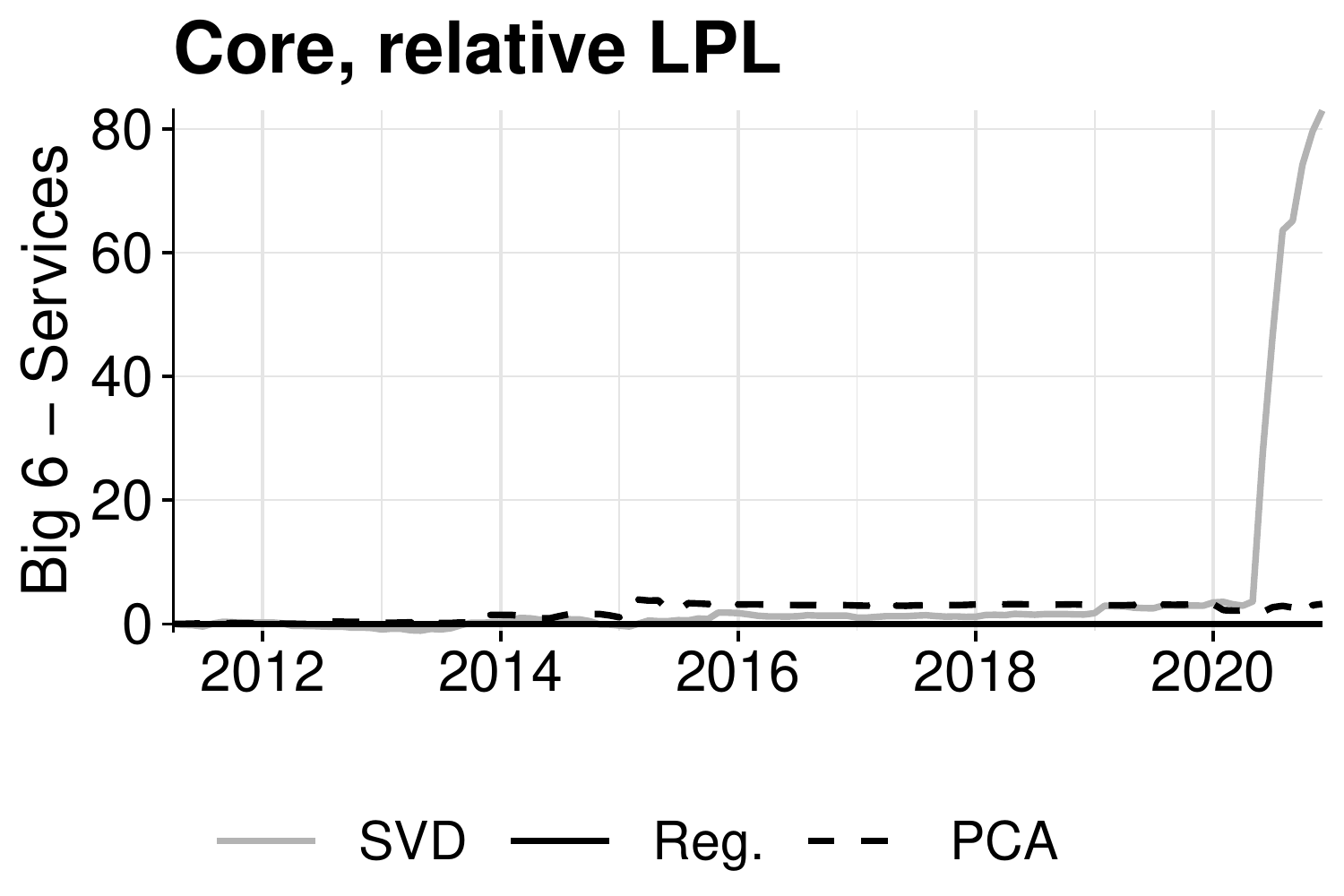}
\includegraphics[width=0.323333333333333\textwidth]{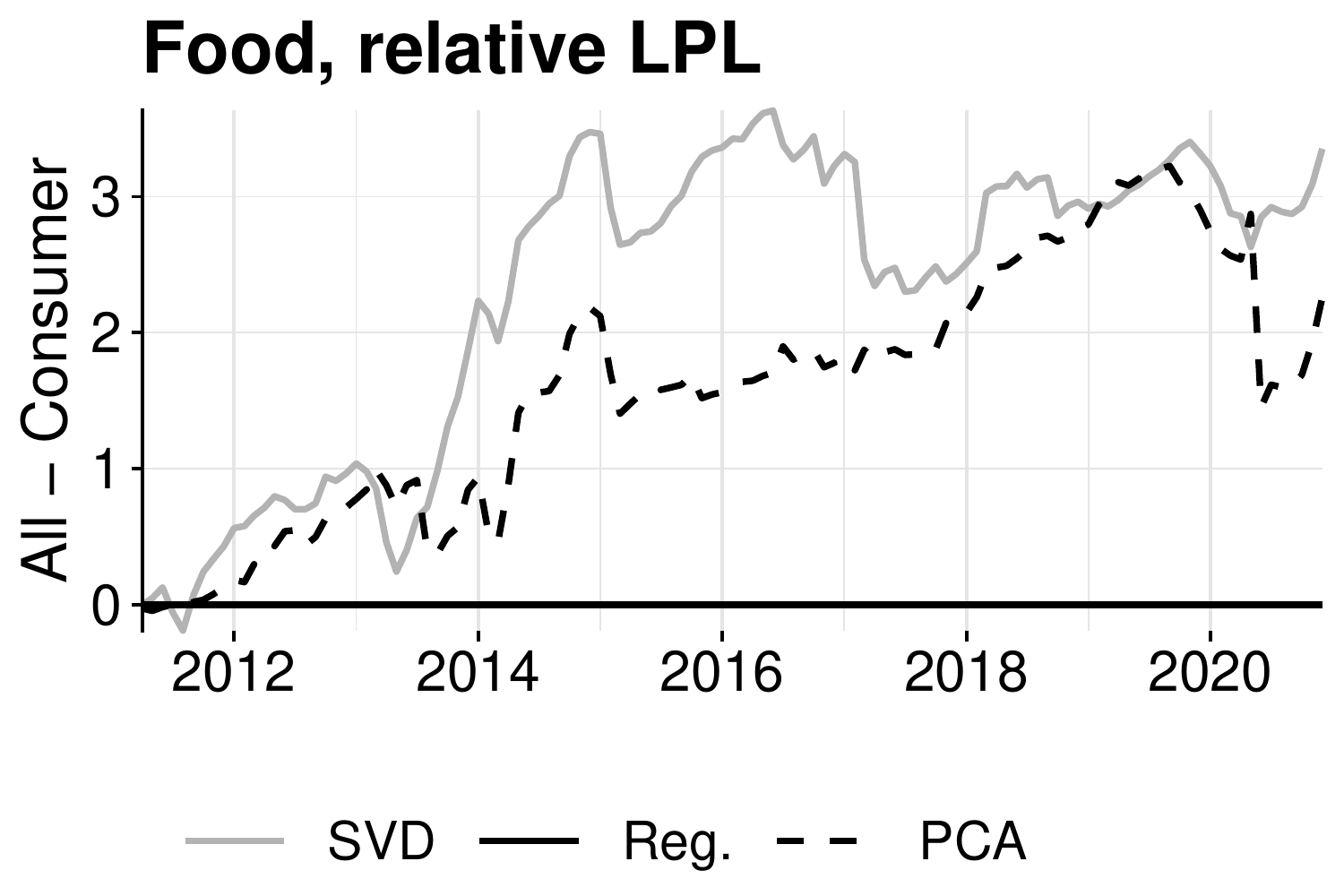}
\includegraphics[width=0.323333333333333\textwidth]{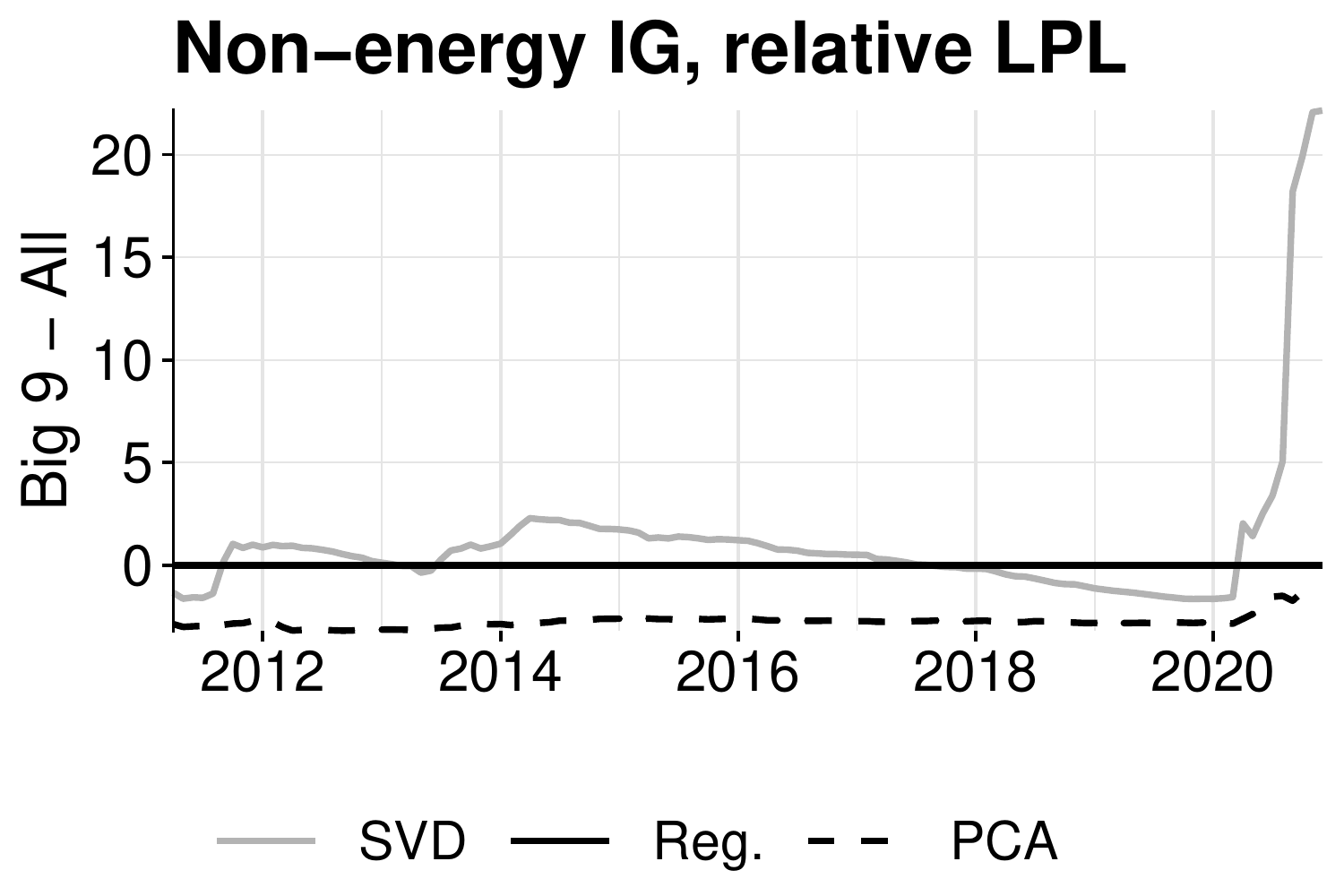}
\includegraphics[width=0.323333333333333\textwidth]{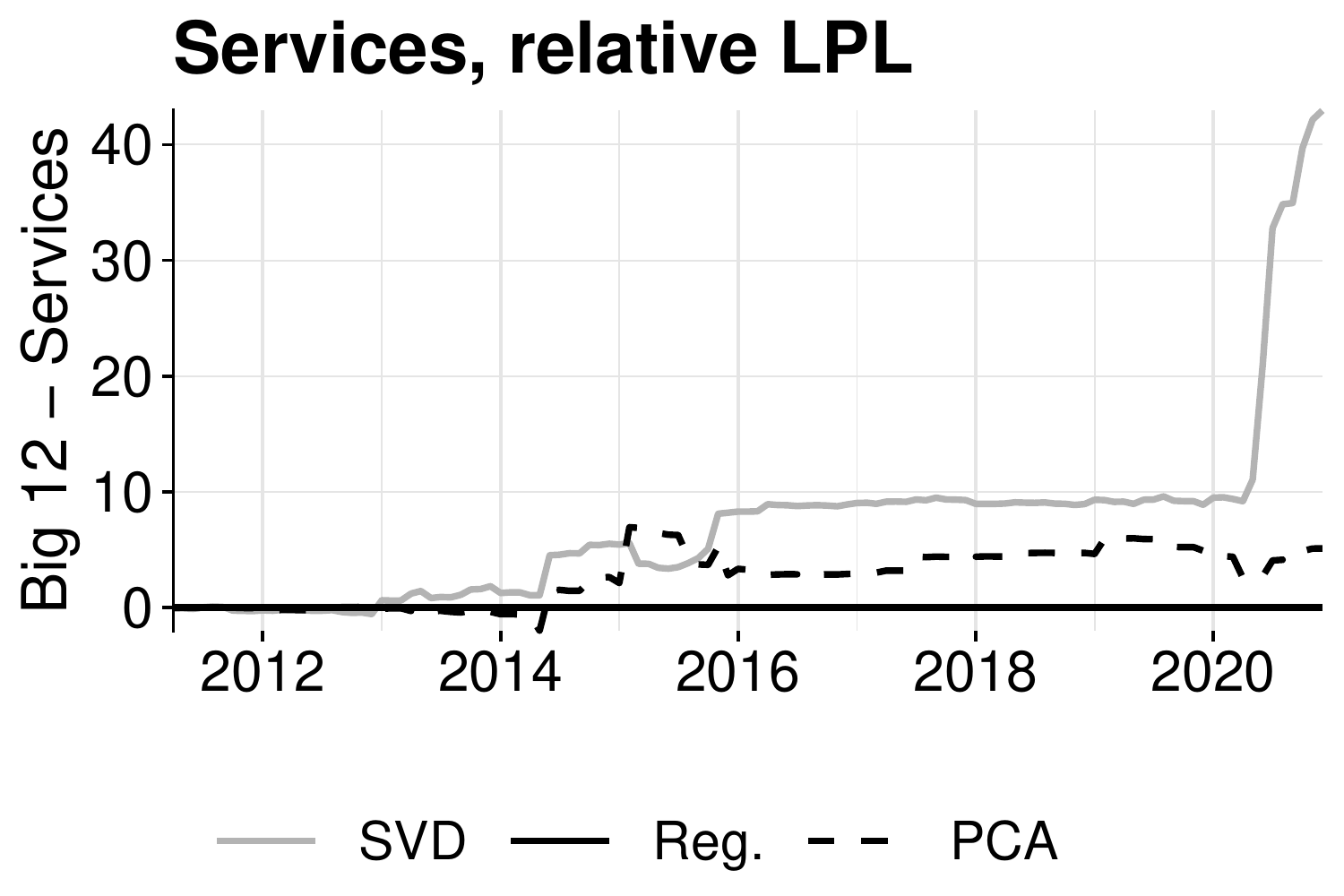}
\includegraphics[width=0.323333333333333\textwidth]{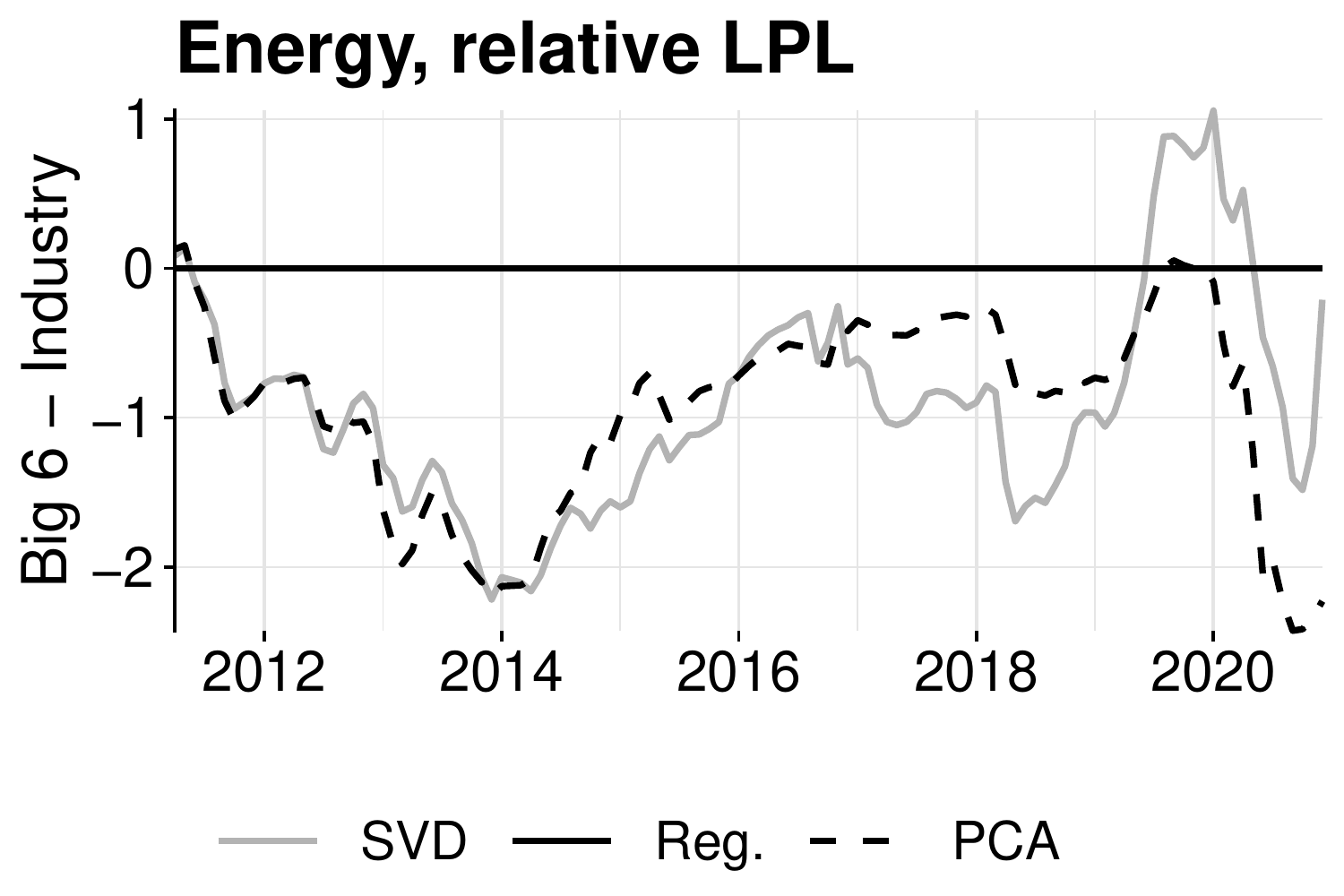}
\includegraphics[width=0.323333333333333\textwidth]{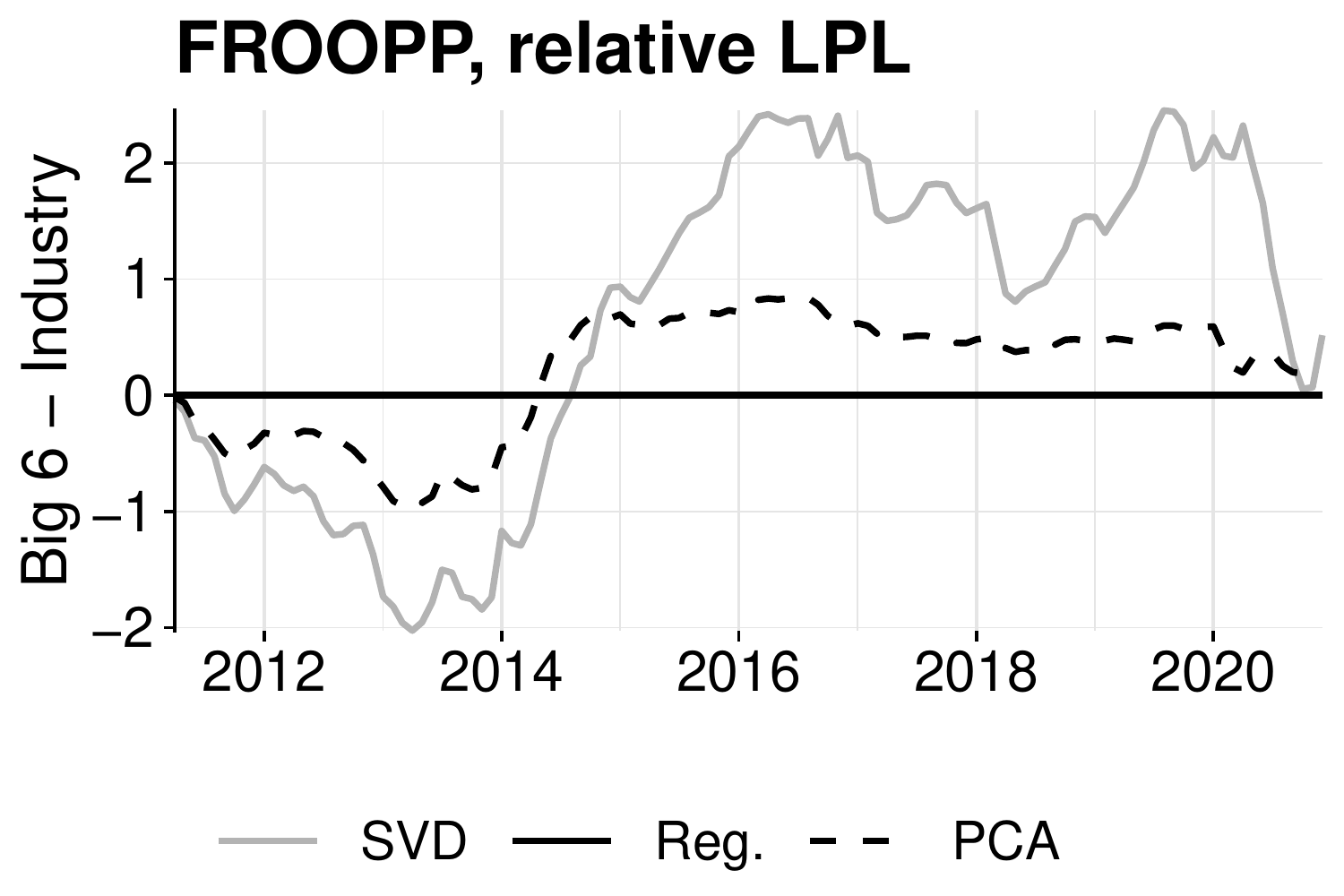}
\caption{Relative LPLs for the best performing SVD model information set by component for 3-step-ahead forecasts.}
\label{fig:lplt_h3}
\end{figure}

In Figure \ref{fig:lplt_h3}, we compare forecast performance over time via the evolution of cumulative log-predictive likelihoods (LPLs). They are shown for the respective best-performing SVD model, alongside the non-survey regression (which serves as the benchmark), and a specification labeled PCA. The PCA specification is constructed in the spirit of \citet{stock2002macroeconomic}. Here, we extract a number of principal components (PCs) from the high-dimensional survey information set, and use the resulting series in our predictive regressions. Adding PCs in an otherwise standard regression is a common way to efficiently summarize big quantities of information, especially when there is the presumption that most variables are driven by a few common economic factors.

When compared with PCA, it appears that SVD has a higher capacity of efficiently exploiting information from the surveys. Overall the improvements for PCA are small, while SVD improves the LPL substantially in several cases. We interpret this result as driven by the fact that SVD is able to use information in a more specific manner. By design, PCs reduce the dimensionality of the regression problem, but this results in losses of information. By outright including the series combined with a shrinkage prior, we can select variables with explanatory power more specifically for the series we intend to forecast.

Importantly, Figure \ref{fig:lplt_h3} also shows that, in most cases, strong relative gains from SVD regressions happen during tumultuous economic periods, when simpler models generally underperform: in 2014 and during the onset of the Covid-19 crisis. These are two structural breaks in the data which had a pronounced impact on expectations and through those the inflation outcomes. In fact, by mid-2014, the EA economy risked moving into deflation and the ECB adopted unconventional monetary policy measures. During this period, we observe relative improvements for most components apart from energy.
The second major improvement in performance of the SVD regression happens around the outbreak of the Covid-19 pandemic. The pandemic was an unprecedented shock, and so were the subsequent public health measures necessary to contain the spread of the disease. Economic policy responses were also rapid and massive, and included very large expansionary fiscal policies to help households and companies through the crisis, together with a large monetary accommodation to provide liquidity and loans to enterprises. This unprecedented mix can hardly be captured (less so in real time) using macro quantities, but expectations implicitly provided such information, resulting in dramatic performance gains during the crisis for most components. In particular, we find sizable gains for core inflation, non-energy industrial goods and services.
 
\subsection{Tail forecast accuracy: downside-risk}
Forecasting extreme events is often as important as targeting the central tendency of key policy relevant series. Consider deviations of inflation around its historical mean, which in well-managed contexts is close to the inflation target of the central bank. Small deviations from the mean may be the result of temporary shocks, or even just noise in the data, and do not provide strong policy-relevant signals. Large deviations, in turn, may be a clearer signal of upcoming policy changes. As a consequence firms, households with loans, and in general economic agents that are affected by interest rate developments are more likely to be interested in tail events than, say, whether inflation is 2.0\% or 1.9\%.

In what follows we concentrate on the lower tail of inflation (corresponding results for the upper tail and for both tails are reported in Appendix). There are two important reasons behind this choice. First, for most of our holdout from 2012 to the end of 2020, inflation has been unexpectedly low and economists speak of a ``missing inflation'' puzzle, whereby models consistently overpredicted inflation. Second, low inflation is of particular interest for monetary policy, as it also defines those values of inflation that can lead monetary policy to the effective lower bound (ELB). Very low inflation would require a more substantial monetary policy stimulus, which however could be problematic given the lower bound on nominal interest rates. The ELB implies an increase in real interest rates and in the cost of capital, lowering firms' investments and economic prospects. Additionally, consumers may be postponing consumption, and the economy can fall in a ``liquidity trap'' where low or negative inflation is self-sustaining. 

\begin{table}[t]
    \centering
    \includegraphics[width=\textwidth]{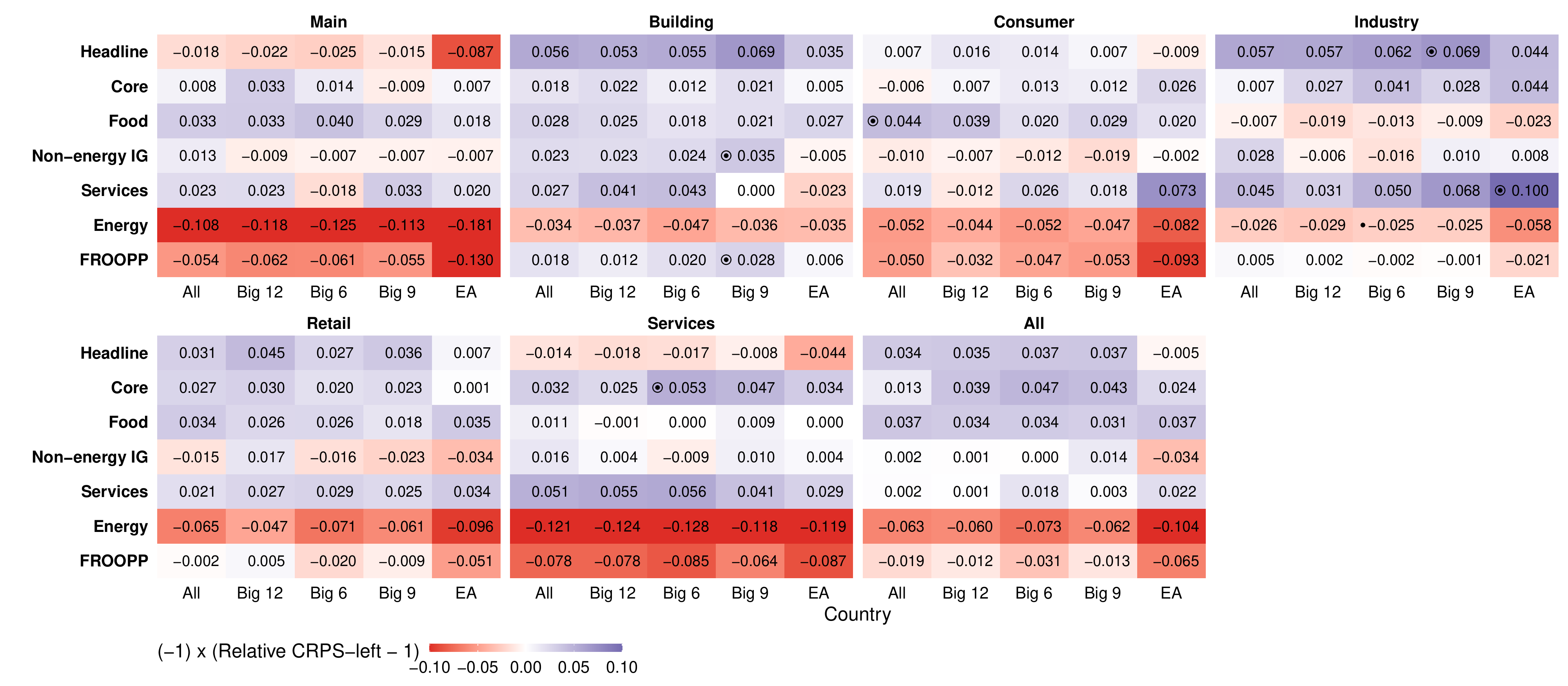}
    \caption{Three-step continuous ranked probability scores (left tail) for models featuring survey expectations relative to the non-survey information set.}
    \label{fig:CRPSlh3}
\end{table}

Figure \ref{fig:CRPSlh3} shows that surveys are helpful in this task for all inflation components except, as for the preceding discussions, energy and FROOPP. For the remaining components, the positive contributions of surveys are comparatively homogeneous, with the notable exception of the main indicators not being useful to predict downside risk in headline inflation. Again, the final block signals that our largest model equipped with the ridge-type prior is capable of extracting signals across all surveys. For headline and core inflation, industry and services surveys provide useful information.

In our final exercise, we chose the two best performing models for forecasting downside-risk in headline and core inflation, and assess their performance across the whole predictive distribution via QSs (in five percent increments of its percentiles). The benchmark is, again, the model without survey information.

\begin{figure}[ht]
    \centering
    \includegraphics[width=1\textwidth]{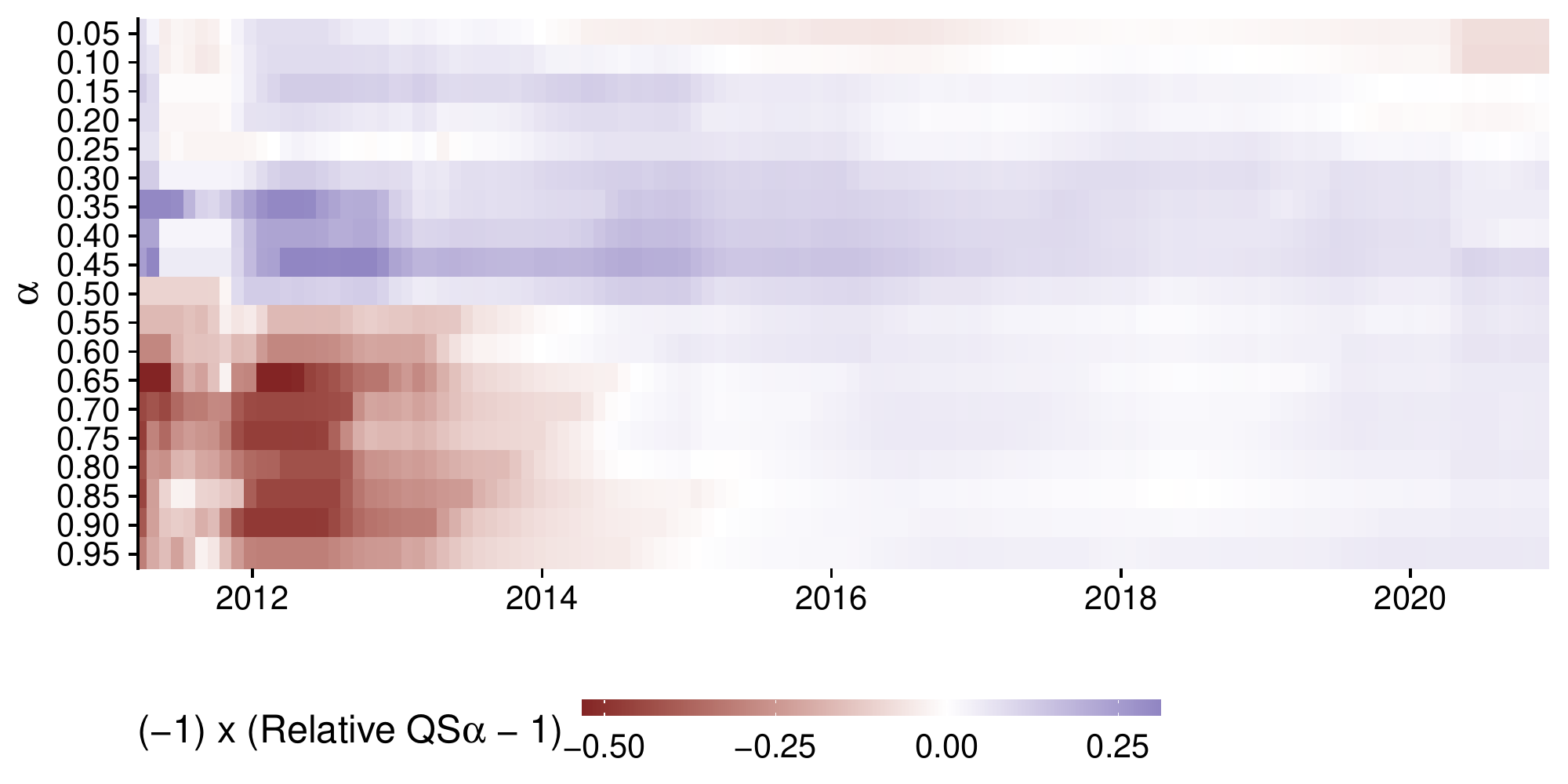}
    \caption{Three-step ahead cumulative quantile scores for headline inflation (survey specification: Industry, Big 9) relative to the benchmark.}
    \label{fig:Headline-Qheatlh3}
\end{figure}
\begin{figure}[ht]
    \centering
    \includegraphics[width=1\textwidth]{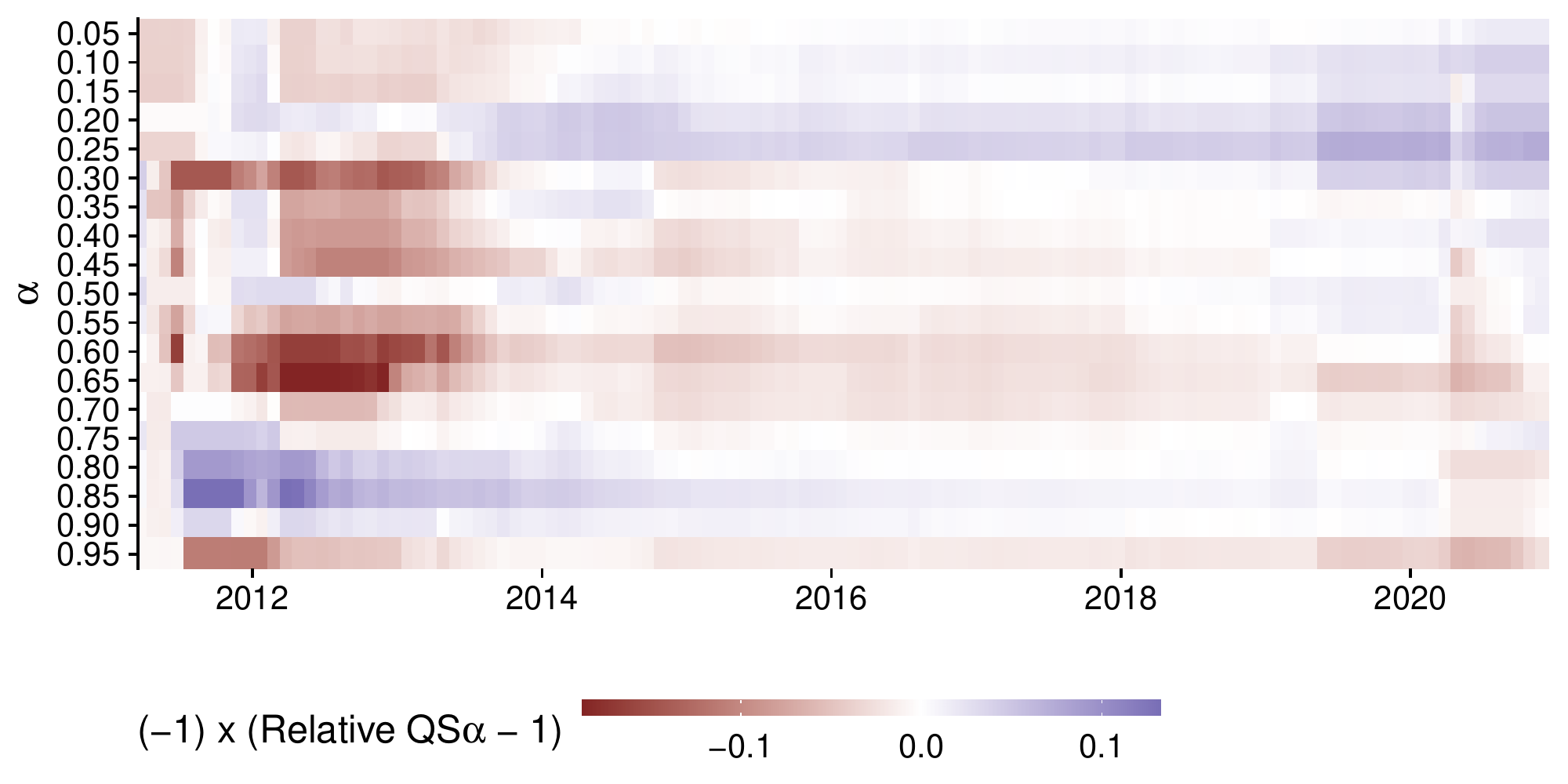}
    \caption{Three-step ahead cumulative quantile scores for core inflation (survey specification: Services, Big 6) relative to the benchmark.}
    \label{fig:Core-qsheath3}
\end{figure}

Figure \ref{fig:Headline-Qheatlh3} refers to headline inflation, and we report the model using surveys of services. For the right tail of the distribution, the survey-model seems to perform worse than the benchmark in the first years of the sample, but improves after 2014. However, surveys provide a clear advantage through the whole sample in the $15$th to $45$th percentiles, suggesting that the model is particularly suited for capturing movements in the lower tail of inflation, and deflationary risks.

Figure \ref{fig:Core-qsheath3} reports the corresponding statistics for core inflation, using industry surveys. The general pattern is similar to the one for headline inflation, with performance improving after 2014. However, we can also observe that the percentiles around the $80$th benefit from survey information in the first part of the sample, when core was decreasing, and lower percentiles around the $20$th thereafter, in correspondence with a slow increase in core inflation. This indicates that the survey model is faster in indicating the direction of movement of the target variable. And it helps in more accurately capturing the dispersion of the distribution of inflation. Moreover, note that the best performance of the model occurs in more recent years, when core inflation started to move more decisively. In this case, the model indeed becomes more useful when it is most needed.

\section{Conclusions}
In this paper, we forecast EA inflation and its components using an econometric model which exploits an enormous number of time series on survey expectations. We focus on the explanatory power of the European Commission's Business and Consumer Survey. Compared to other surveys, this survey is timely (its frequency is monthly), covers a wide population of firms and consumers, and includes question about recent developments as well as intentions of future behaviour. Focusing on the forecast of inflation, we highlight the role of these expectations and their overall contribution to inflation movements. 

In the largest case, the resulting regression features about 1400 predictors. To make estimation tractable, we provide an efficient sampling algorithm based on a ridge-type prior for predictive inference. We find that survey information improves forecasts of headline inflation and its components in many cases, both for point, density and tail forecasts. Regressions featuring solely aggregates of surveys, or simple factor-based methods to compress high-dimensional information, are typically outperformed by our proposed approach of outright including all information. We detect pronounced relative gains, particularly for metrics capturing higher-order moments of the predictive distribution, during unusual economic periods such as the European sovereign debt crisis and the onset of the Covid-19 pandemic.

The implications for policymakers are straightforward. First, we confirm the well-know result that expectations are a major determinant of price dynamics, and lots of attention should be given to their measurement. Second, we show that useful information can be drawn by surveys that are timely, broad in audience and scope, and both forward- and backward-looking.

\small{\setstretch{0.85}
\addcontentsline{toc}{section}{References}
\bibliographystyle{frbcle.bst}
\bibliography{lit}}\normalsize\clearpage

@TechReport{Banbura_etal21,
  author={Bańbura, Marta and Leiva-Leon, Danilo and Menz, Jan-Oliver},
  title={{Do inflation expectations improve model-based inflation forecasts?}},
  year=2021,
  month=Oct,
  institution={European Central Bank},
  type={Working Paper Series},
  url={https://ideas.repec.org/p/ecb/ecbwps/20212604.html},
  number={2604},
  doi={},
}

@TechReport{Moretti_etal19,
  author={Moretti, Laura and Onorante, Luca and Zakipour Saber, Shayan},
  title={{Phillips curves in the euro area}},
  year=2019,
  month=Jul,
  institution={European Central Bank},
  type={Working Paper Series},
  url={https://ideas.repec.org/p/ecb/ecbwps/20192295.html},
  number={2295},
  doi={},
}

@Article{ECB_SPF,
  author={{De Vincent-Humphreys}, Rupert and Dimitrova, Ivelina and Falck, Elisabeth and Henkel, Lukas and Meyler, Aidan},
  title={{Twenty years of the ECB Survey of Professional Forecasters}},
  journal={Economic Bulletin Articles},
  year=2019,
  volume={1},
  number={},
  pages={},
  month={},
  doi={},
  url={https://ideas.repec.org/a/ecb/ecbart/201900011.html}
}

@TechReport{Arioli17,
  author={Arioli, Rodolfo and Bates, Colm and Dieden, Heinz Christian and Duca, Ioana and Friz, Roberta and Gayer, Christian and Kenny, Geoff and Meyler, Aidan and Pavlova, Iskra},
  title={{EU consumers’ quantitative inflation perceptions and expectations: an evaluation}},
  year=2017,
  month=Apr,
  institution={European Central Bank},
  type={Occasional Paper Series},
  url={https://ideas.repec.org/p/ecb/ecbops/2017186.html},
  number={186},
  doi={},
}

@article{geweke2010comparing,
  title={Comparing and evaluating Bayesian predictive distributions of asset returns},
  author={Geweke, John and Amisano, Gianni},
  journal={International Journal of Forecasting},
  volume={26},
  number={2},
  pages={216--230},
  year={2010},
  publisher={Elsevier}
}

@article{gneiting2011comparing,
  title={Comparing density forecasts using threshold-and quantile-weighted scoring rules},
  author={Gneiting, Tilmann and Ranjan, Roopesh},
  journal={Journal of Business \& Economic Statistics},
  volume={29},
  number={3},
  pages={411--422},
  year={2011},
  publisher={Taylor \& Francis}
}

@article{stock2002macroeconomic,
  title={Macroeconomic forecasting using diffusion indexes},
  author={Stock, James H and Watson, Mark W},
  journal={Journal of Business \& Economic Statistics},
  volume={20},
  number={2},
  pages={147--162},
  year={2002},
  publisher={Taylor \& Francis}
}

@article{hauzenberger2021fast,
  title={Fast and flexible Bayesian inference in time-varying parameter regression models},
  author={Hauzenberger, Niko and Huber, Florian and Koop, Gary and Onorante, Luca},
  journal={Journal of Business \& Economic Statistics},
  pages={1--15},
  year={2021},
  publisher={Taylor \& Francis}
}

@inproceedings{trippe2019lr,
  title={LR-GLM: High-dimensional Bayesian inference using low-rank data approximations},
  author={Trippe, Brian and Huggins, Jonathan and Agrawal, Raj and Broderick, Tamara},
  booktitle={International Conference on Machine Learning},
  pages={6315--6324},
  year={2019},
  organization={PMLR}
}

@article{ciccarelli2021expectation,
  title={Expectation spillovers and the return of inflation},
  author={Ciccarelli, Matteo and Garc{\'\i}a, Juan Angel},
  journal={Economics Letters},
  volume={209},
  pages={110119},
  year={2021},
  publisher={Elsevier}
}

@article{carvalho2010horseshoe,
  title={The horseshoe estimator for sparse signals},
  author={Carvalho, Carlos M and Polson, Nicholas G and Scott, James G},
  journal={Biometrika},
  volume={97},
  number={2},
  pages={465--480},
  year={2010},
  publisher={Oxford University Press}
}

@TechReport{ClaveriaAl21,
  author={Oscar Claveria and Enric Monte and Salvador Torra},
  title={{\&quot;Nowcasting and forecasting GDP growth with machine-learning sentiment indicators\&quot;}},
  year=2021,
  month=Feb,
  institution={University of Barcelona, Research Institute of Applied Economics},
  type={IREA Working Papers},
  url={https://ideas.repec.org/p/ira/wpaper/202103.html},
  number={202103},
  doi={},
}

@TechReport{ClaveriaAl20,
  author={Oscar Claveria and Enric Monte and Salvador Torra},
  title={{“Spectral analysis of business and consumer survey data”}},
  year=2020,
  month=May,
  institution={University of Barcelona, Regional Quantitative Analysis Group},
  type={AQR Working Papers},
  url={https://ideas.repec.org/p/aqr/wpaper/202002.html},
  number={2012002},
  doi={},
}

@Article{ClaveriaAl07,
  author={Claveria, Oscar and Pons, Ernest and Ramos, Raul},
  title={{Business and consumer expectations and macroeconomic forecasts}},
  journal={International Journal of Forecasting},
  year=2007,
  volume={23},
  number={1},
  pages={47-69},
  month={},
  doi={},
  url={https://ideas.repec.org/a/eee/intfor/v23y2007i1p47-69.html}
}

@TechReport{ECBconsSurvey20,
  author={Bańnkowska, Katarzyna and Borlescu, Ana Maria and Charalambakis, Evangelos and Da Silva, António Dias and Di Laurea, Davide and Dossche, Maarten and Georgarakos, Dimitris and Honkkila, Juha and Kennedy, Neale and Kenny},
  title={{ECB Consumer Expectations Survey: an overview and first evaluation}},
  year=2021,
  month=Dec,
  institution={European Central Bank},
  type={Occasional Paper Series},
  url={https://ideas.repec.org/p/ecb/ecbops/2021287.html},
  number={287},
  doi={},
}

@Article{GiannoneetAl14,
  author={Giannone, Domenico and Lenza, Michele and Momferatou, Daphne and Onorante, Luca},
  title={{Short-term inflation projections: A Bayesian vector autoregressive approach}},
  journal={International Journal of Forecasting},
  year=2014,
  volume={30},
  number={3},
  pages={635-644},
  month={},
  doi={10.1016/j.ijforecast.2013},
  url={https://ideas.repec.org/a/eee/intfor/v30y2014i3p635-644.html}
}

@TechReport{DelNegroetAl20,
  author={Del Negro, Marco and Lenza, Michele and Primiceri, Giorgio E. and Tambalotti, Andrea},
  title={{What’s up with the Phillips Curve?}},
  year=2020,
  month=Jul,
  institution={European Central Bank},
  type={Working Paper Series},
  url={https://ideas.repec.org/p/ecb/ecbwps/20202435.html},
  number={2435},
  doi={},
}

@TechReport{KontogeorgosLmbrias19,
  author={Kontogeorgos, Georgios and Lambrias, Kyriacos},
  title={{An analysis of the Eurosystem/ECB projections}},
  year=2019,
  month=Jun,
  institution={European Central Bank},
  type={Working Paper Series},
  url={https://ideas.repec.org/p/ecb/ecbwps/20192291.html},
  number={2291},
  doi={},
}

@TechReport{AlessietAl14,
  author={Alessi, Lucia and Ghysels, Eric and Onorante, Luca and Potter, Simon and Peach, Richard},
  title={{Central bank macroeconomic forecasting during the global financial crisis: the European Central Bank and Federal Reserve Bank of New York experiences}},
  year=2014,
  month=Jul,
  institution={European Central Bank},
  type={Working Paper Series},
  url={https://ideas.repec.org/p/ecb/ecbwps/20141688.html},
  number={1688},
  doi={},
}

@article{LenzaJarocinski16,
  title={An inflation-predicting measure of the output gap in the euro area},
  author={Jaroci{\'n}ski, Marek and Lenza, Michele},
  journal={Journal of Money, Credit and Banking},
  volume={50},
  number={6},
  pages={1189--1224},
  year={2018},
  publisher={Wiley Online Library}
}

@article{giannone2021economic,
  title={Economic predictions with big data: The illusion of sparsity},
  author={Giannone, Domenico and Lenza, Michele and Primiceri, Giorgio E},
    journal={Econometrica},
  year={2021},
  publisher={Wiley}
}

@article{cong2017fast,
  title={Fast simulation of hyperplane-truncated multivariate normal distributions},
  author={Cong, Yulai and Chen, Bo and Zhou, Mingyuan},
  journal={Bayesian Analysis},
  volume={12},
  number={4},
  pages={1017--1037},
  year={2017},
  publisher={International Society for Bayesian Analysis}
}

@article{griffin2013some,
  title={Some priors for sparse regression modelling},
  author={Griffin, Jim E and Brown, Philip J},
  journal={Bayesian Analysis},
  volume={8},
  number={3},
  pages={691--702},
  year={2013},
  publisher={International Society for Bayesian Analysis}
}

@book{koop2007bayesian,
  title={Bayesian econometric methods},
  author={Koop, Gary and Poirier, Dale J and Tobias, Justin L},
  year={2007},
  publisher={Cambridge University Press}
}


\normalsize\clearpage
\setcounter{page}{1}
\begin{appendices}
\section{One-step ahead forecasts}
\begin{figure}[ht]
    \centering
    \includegraphics[width=\textwidth]{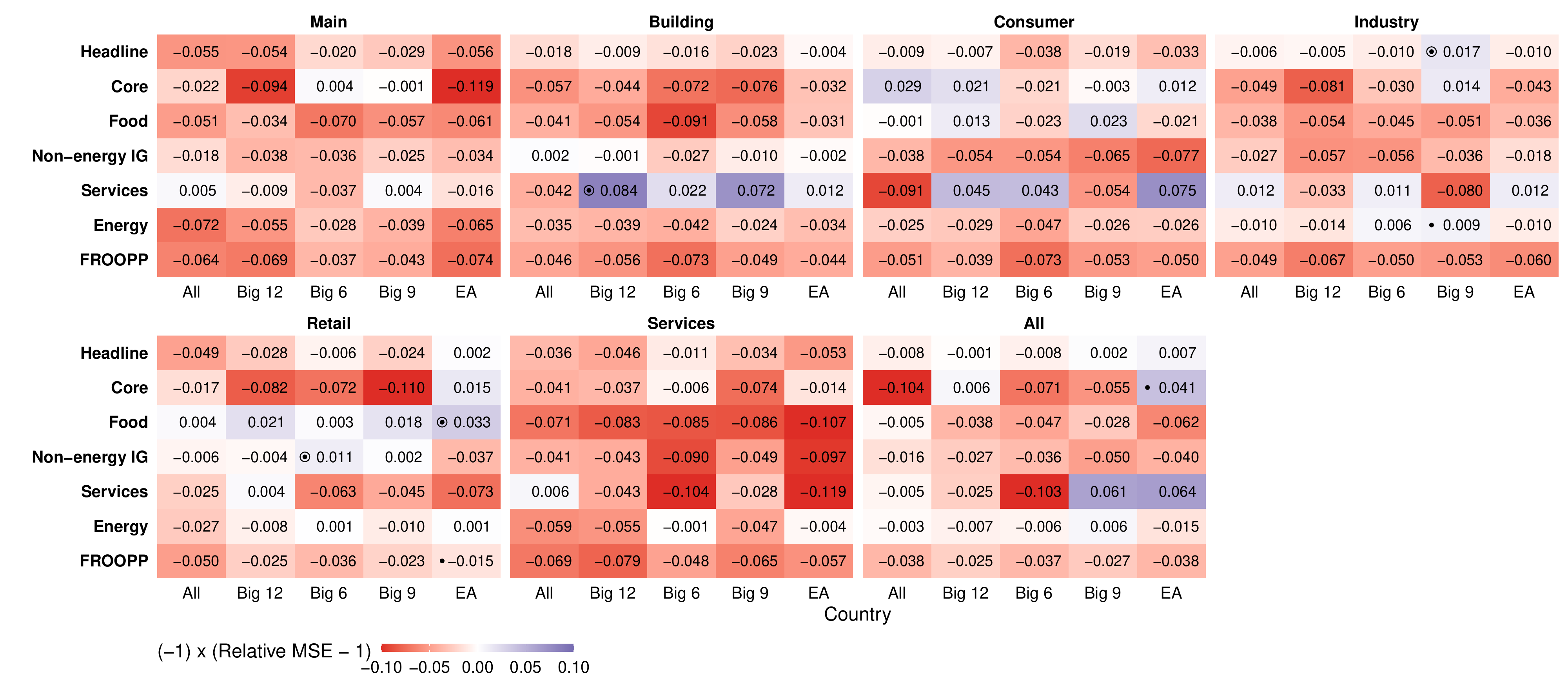}
    \caption{One-step mean squared error for models featuring survey expectations relative to the non-survey information set.}
    \label{fig:MSEh1}
\end{figure}
\begin{figure}[ht]
    \centering
    \includegraphics[width=\textwidth]{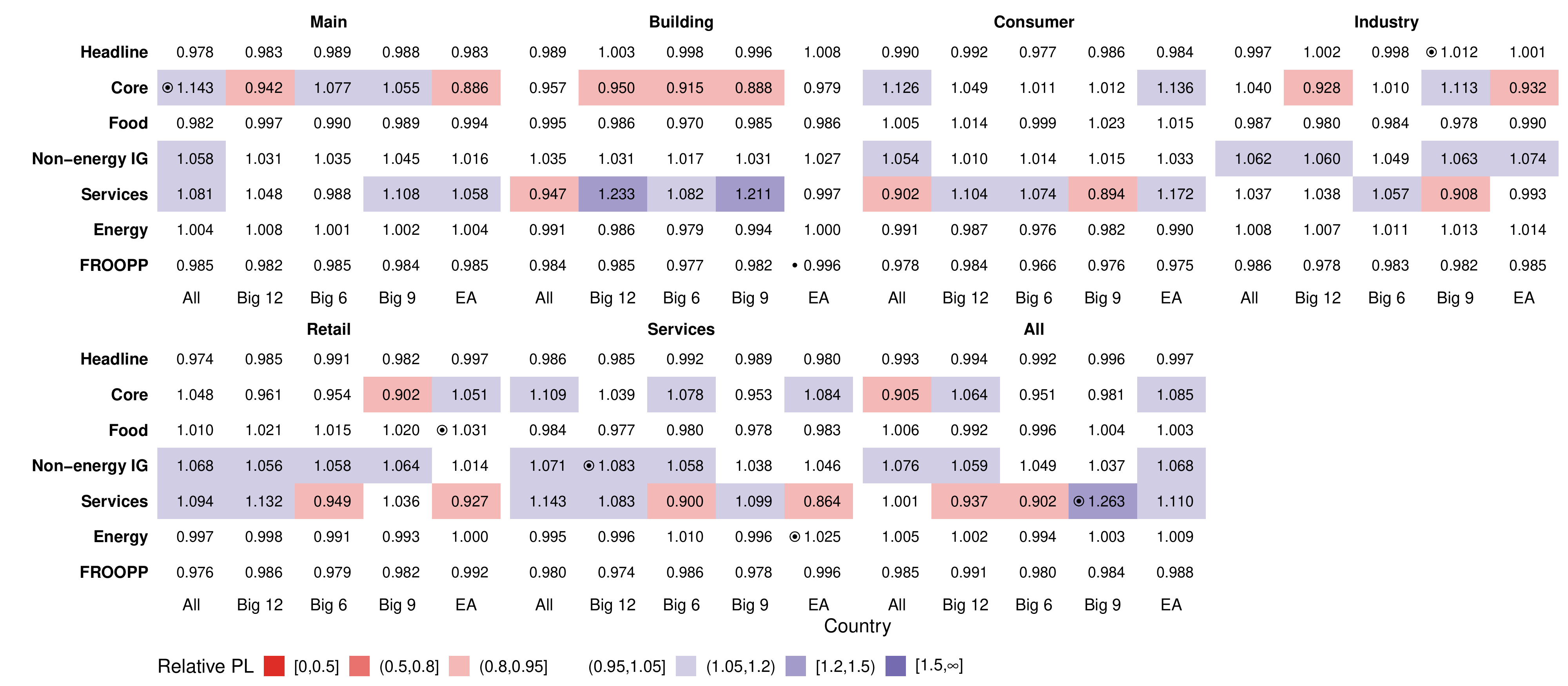}
    \caption{One-step ahead predictive likelihood for models featuring survey expectations relative to the non-survey information set.}
    \label{fig:expLPLh1}
\end{figure}

\begin{figure}[t]
\includegraphics[width=0.323333333333333\textwidth]{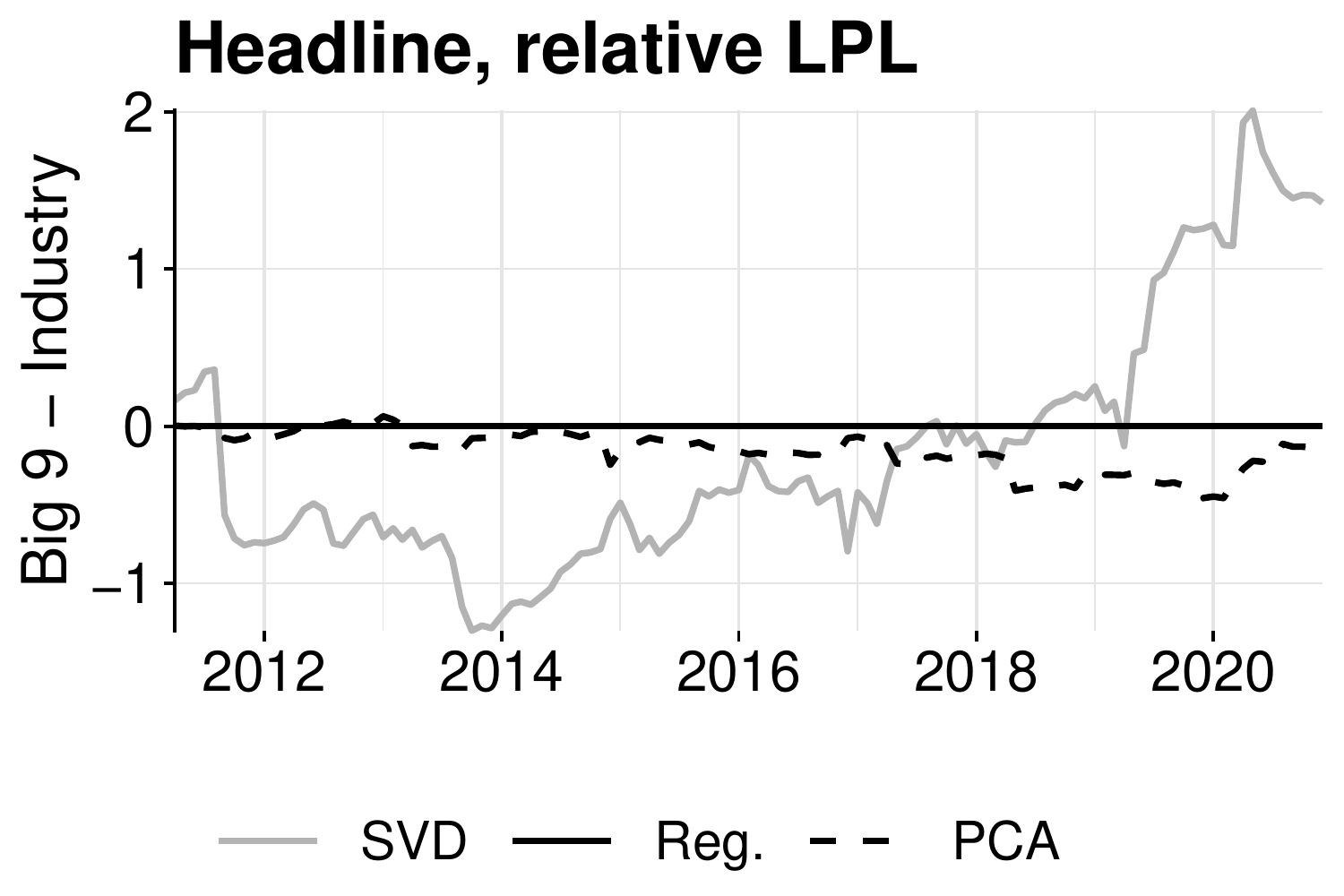}
\includegraphics[width=0.323333333333333\textwidth]{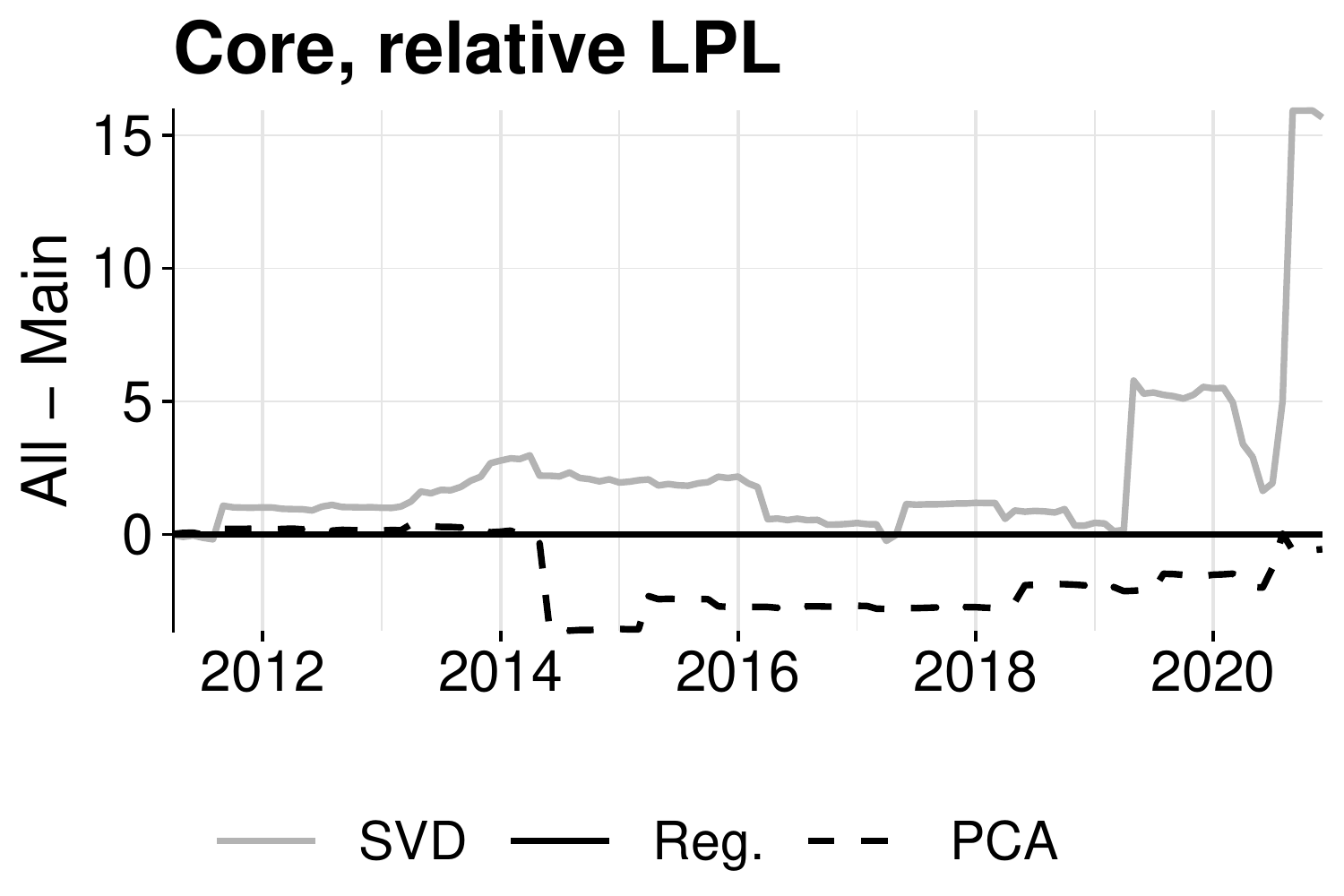}
\includegraphics[width=0.323333333333333\textwidth]{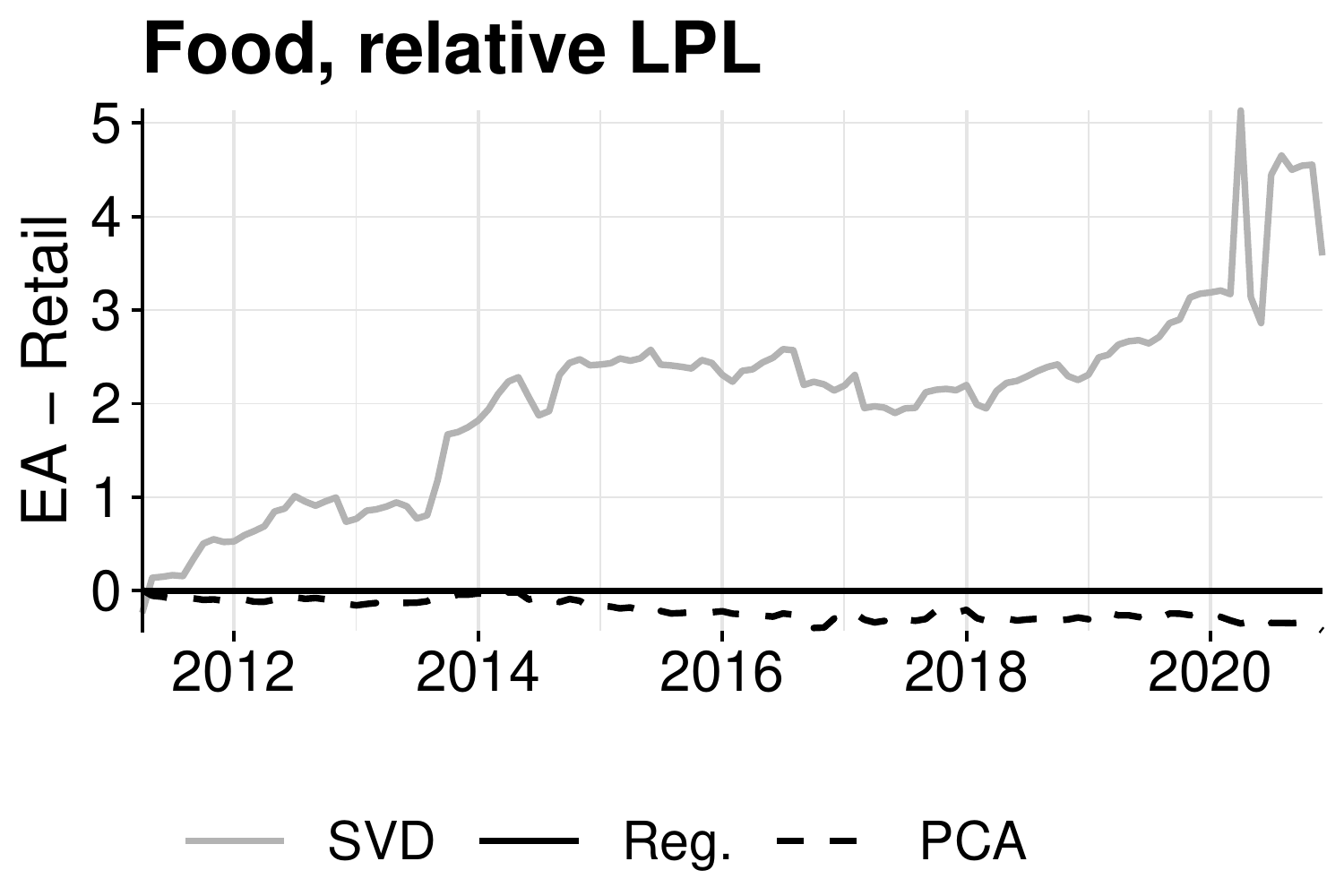}
\includegraphics[width=0.323333333333333\textwidth]{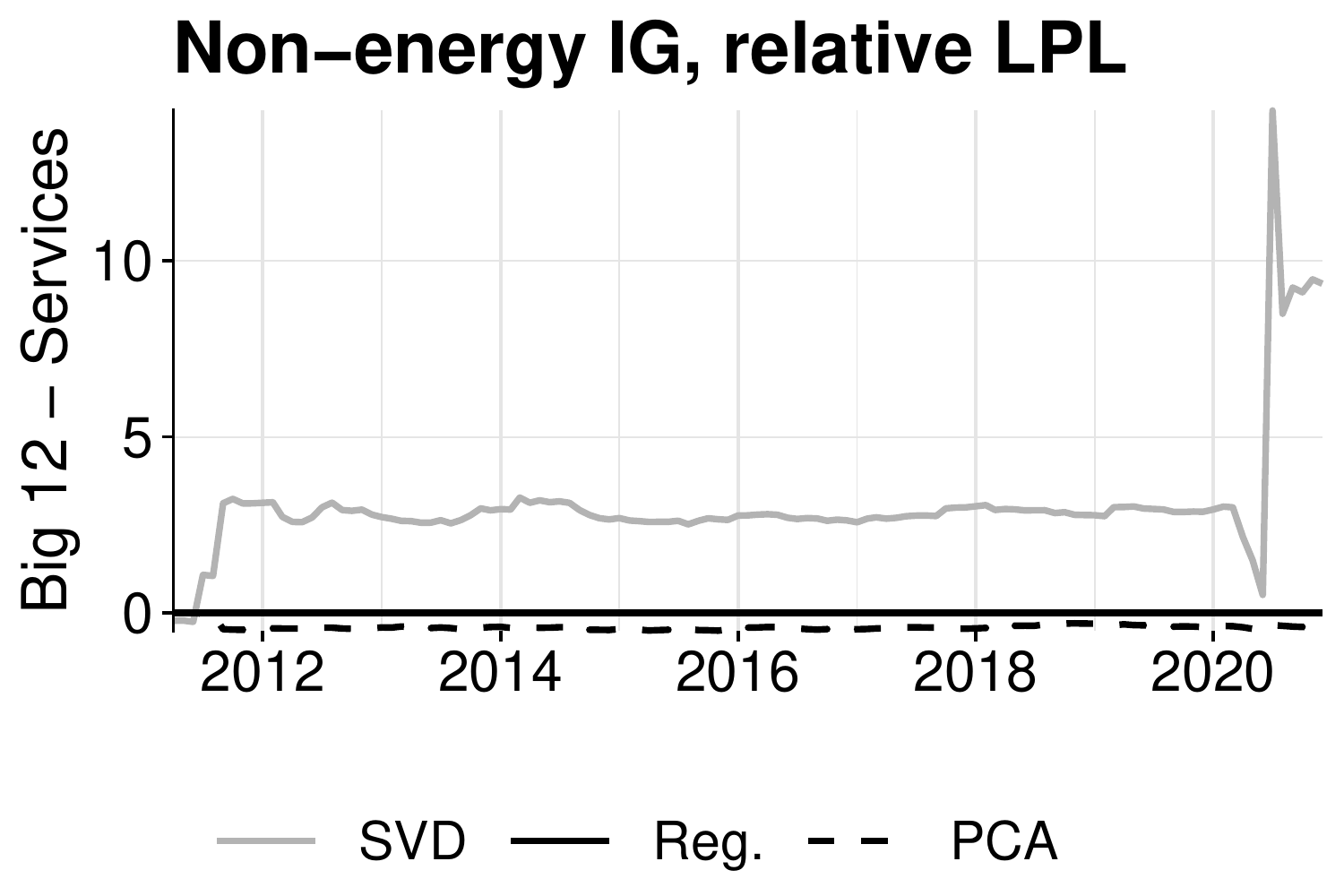}
\includegraphics[width=0.323333333333333\textwidth]{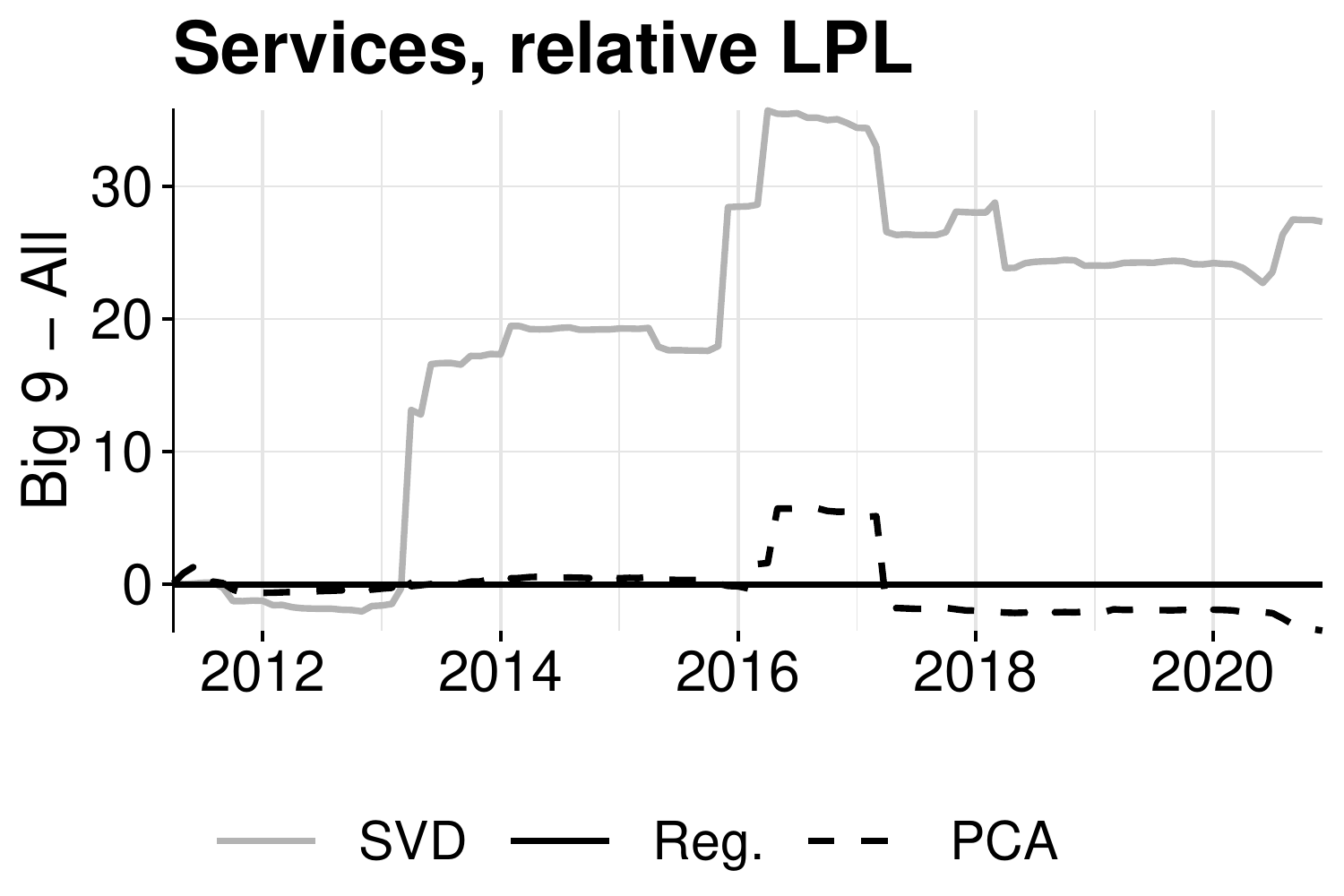}
\includegraphics[width=0.323333333333333\textwidth]{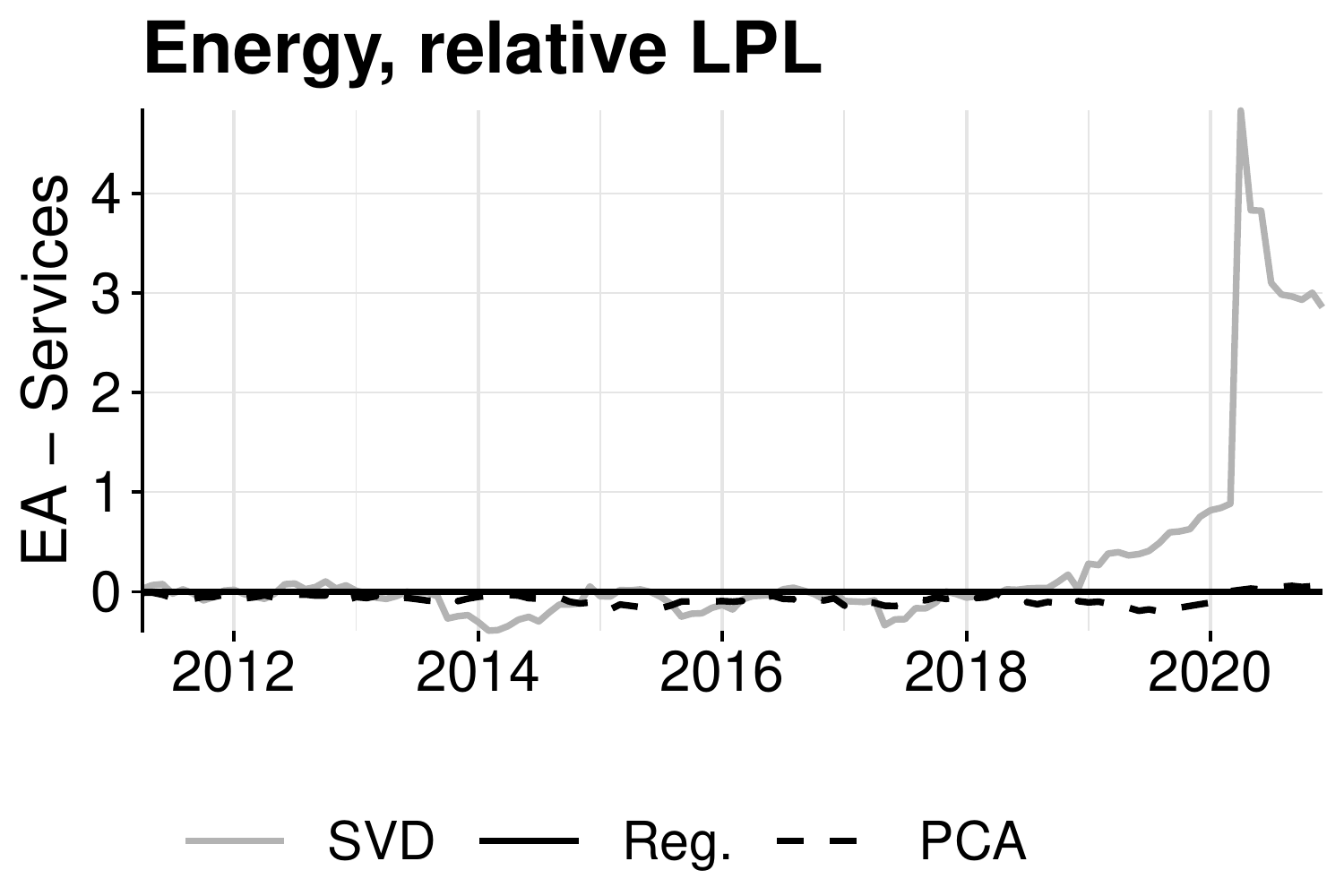}
\includegraphics[width=0.323333333333333\textwidth]{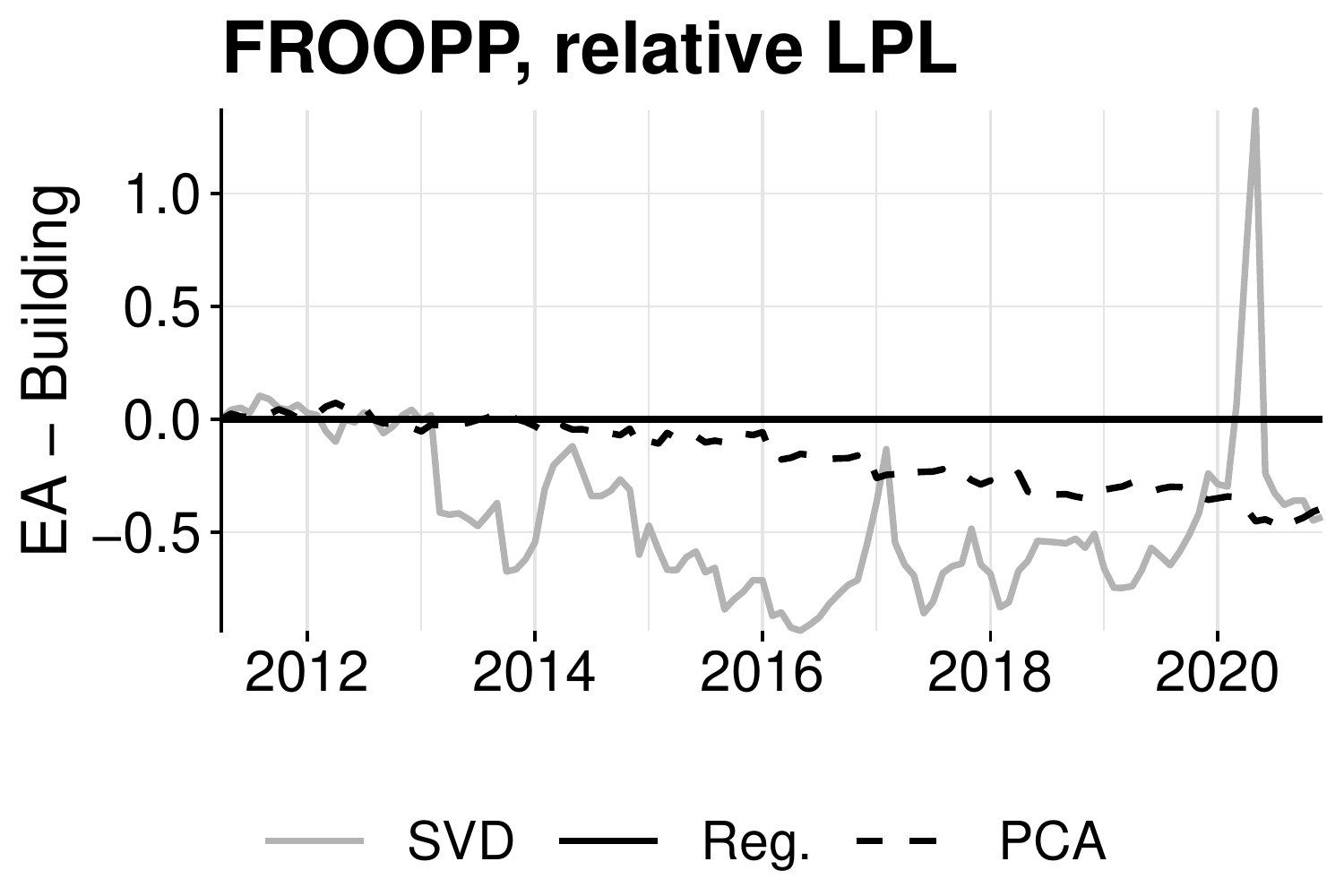}
\caption{Relative LPLs for the best performing SVD model information set by component for 1-step-ahead forecasts.}
\label{fig:lplt_h1}
\end{figure}

\clearpage
\section{Tail forecasts}
\begin{figure}[ht]
    \centering
    \includegraphics[width=\textwidth]{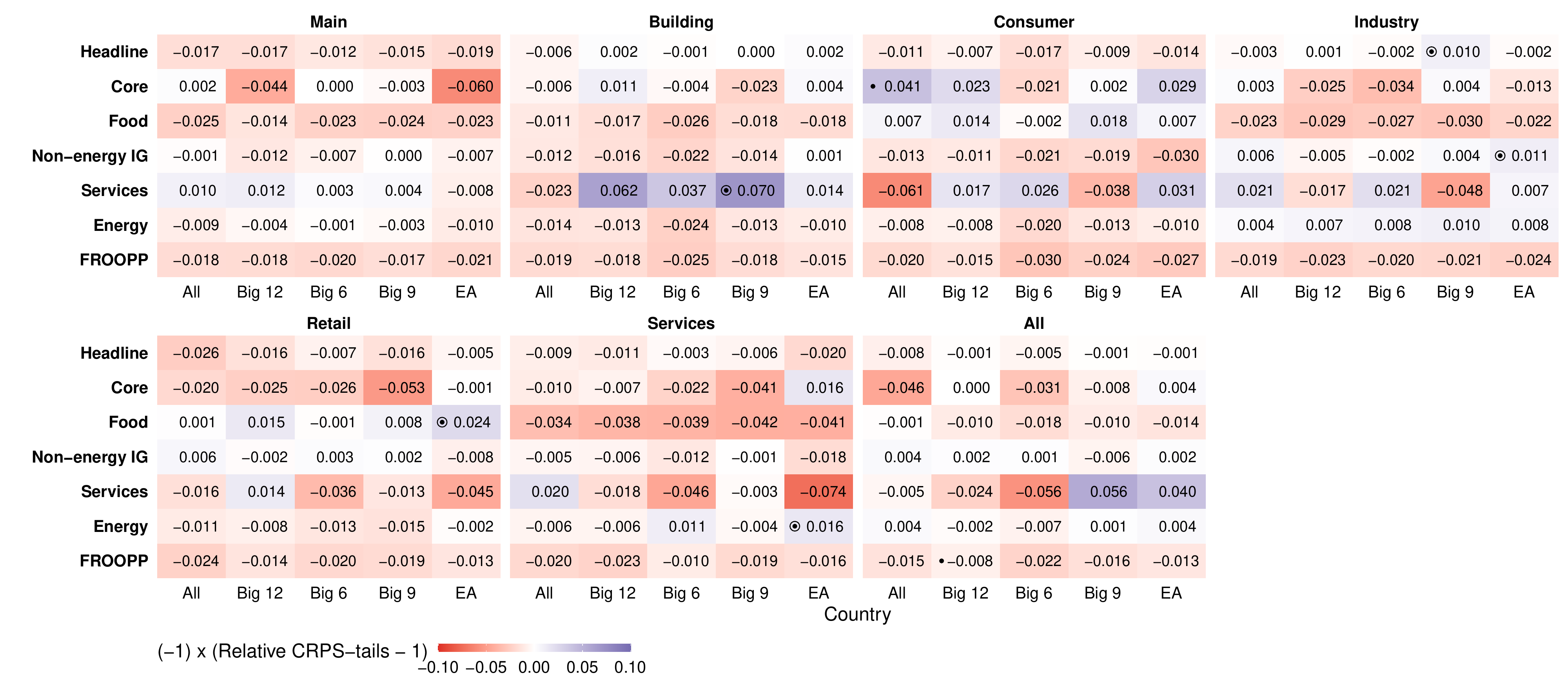}
    \caption{One-step ahead continuous ranked probability scores (both tails) for models featuring survey expectations relative to the non-survey information set.}
    \label{fig:CRPSlh1}
\end{figure}
\begin{figure}[ht]
    \centering
    \includegraphics[width=\textwidth]{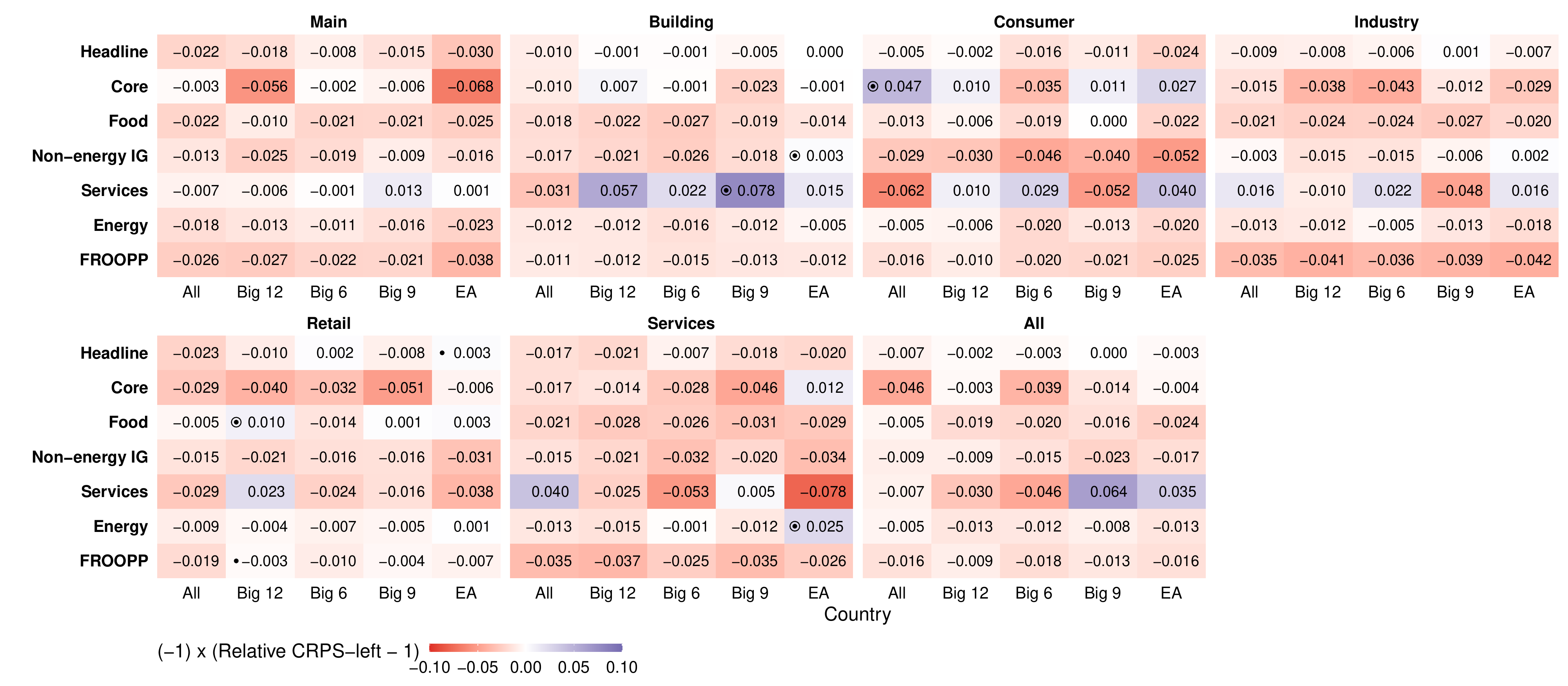}
    \caption{One-step ahead continuous ranked probability scores (left tail) for models featuring survey expectations relative to the non-survey information set.}
    \label{fig:CRPSlh1}
\end{figure}
\begin{figure}[ht]
    \centering
    \includegraphics[width=\textwidth]{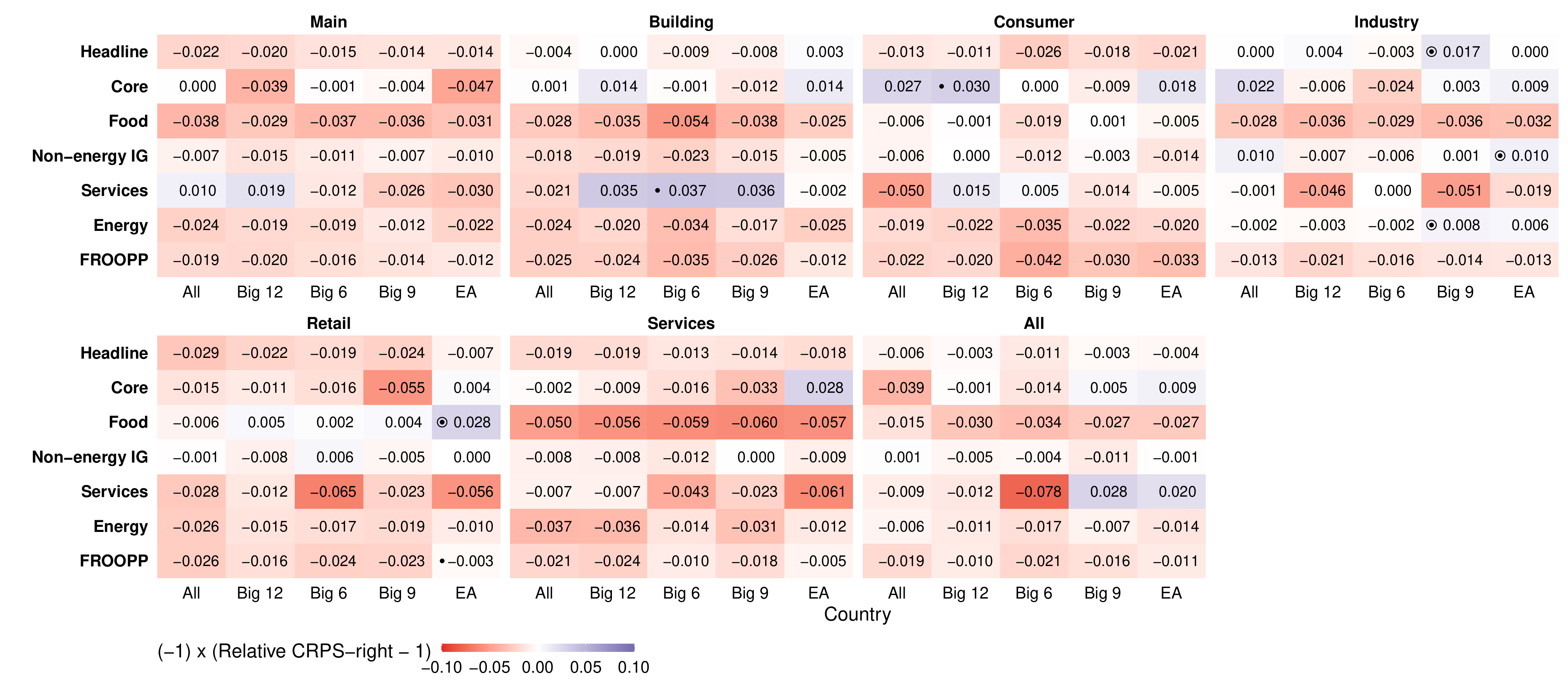}
    \caption{One-step ahead continuous ranked probability scores (right tail) for models featuring survey expectations relative to the non-survey information set.}
    \label{fig:CRPSrh1}
\end{figure}

\begin{figure}[ht]
    \centering
    \includegraphics[width=\textwidth]{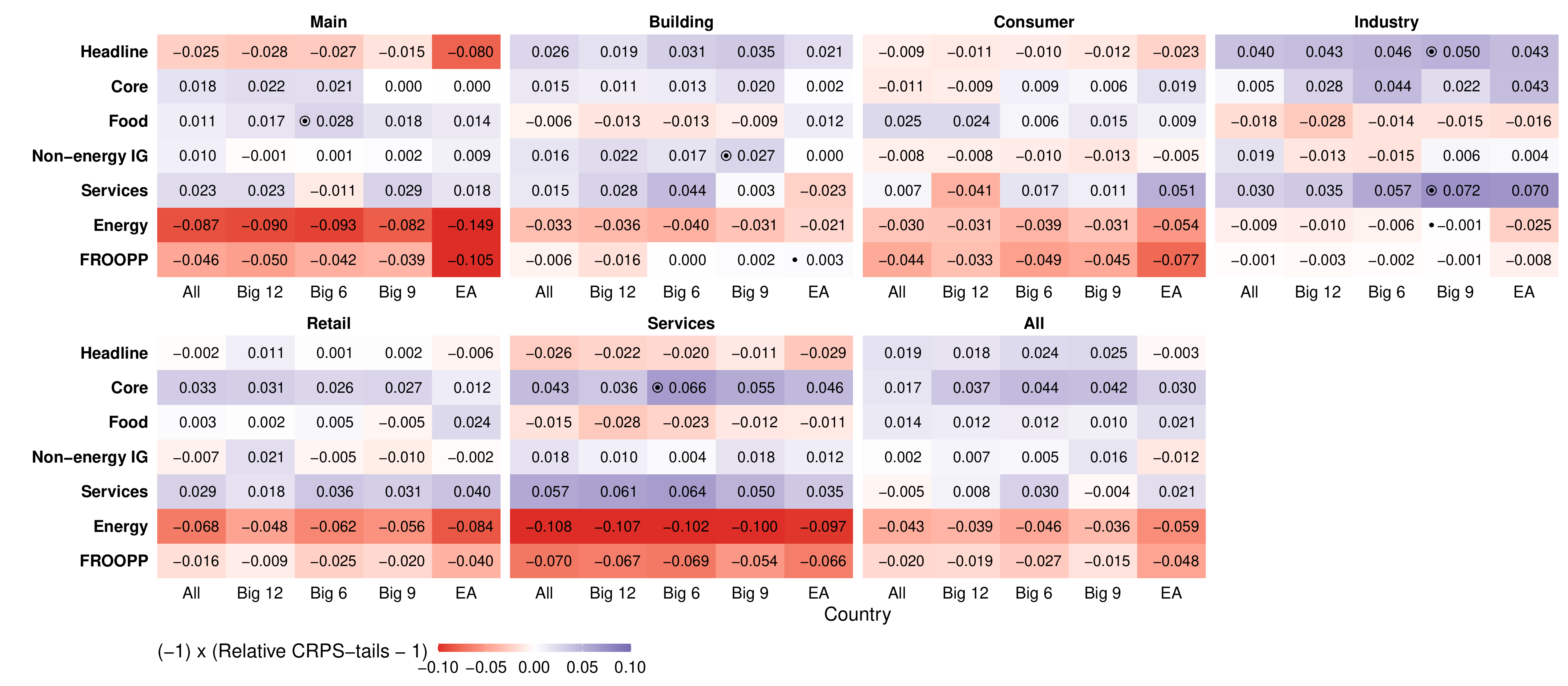}
    \caption{Three-step ahead continuous ranked probability scores (both tails) for models featuring survey expectations relative to the non-survey information set.}
    \label{fig:CRPSrh1}
\end{figure}

\begin{figure}[ht]
    \centering
    \includegraphics[width=\textwidth]{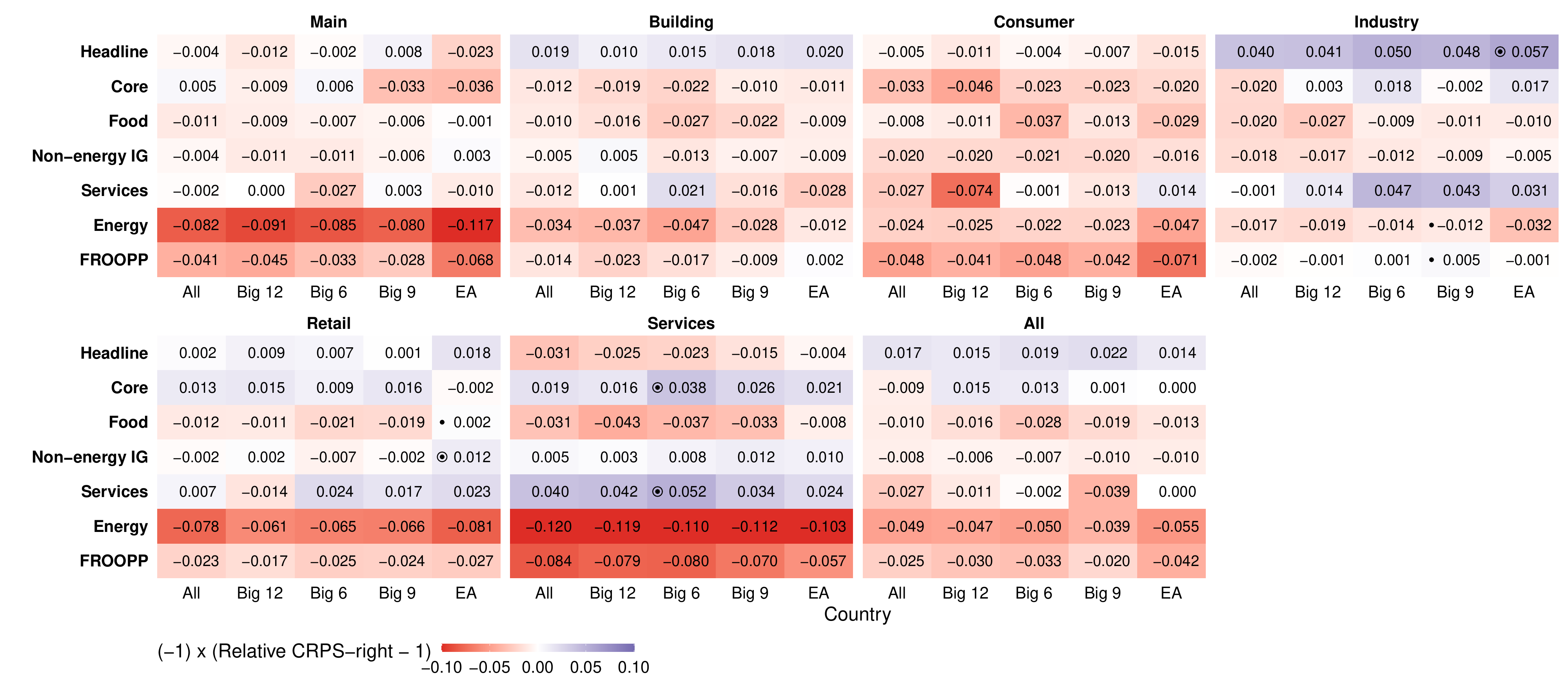}
    \caption{Three-step ahead continuous ranked probability scores (right tail) for models featuring survey expectations relative to the non-survey information set.}
    \label{fig:CRPSrh1}
\end{figure}

\end{appendices}
\end{document}